\documentclass{aa}
\usepackage{booktabs}
\usepackage{graphicx}
\usepackage{txfonts}
\usepackage{lipsum}
\usepackage{subcaption}  
\usepackage{xcolor}

\usepackage{lscape}             % to rotate a single page table, example in appendix.
\usepackage{placeins}           % useful with \FloatBarrier, to keep 
\usepackage{hyperref}
\hypersetup{linkcolor=blue,citecolor=blue,colorlinks=true,pdfborderstyle={}}
\begin{document}

   \title{Self-consistent modeling of energy-dependent synchrotron polarization in blazars with application to Mrk 421}

   \author{F. Apel
          \inst{\ref{rub}}
          \and
          M. Tisang\inst{\ref{nwu}}
          \and
          V. Barbosa Martins\inst{\ref{ufg}, \ref{rub}}
          \and
          X. Rodrigues\inst{\ref{apc}}
          \and
          M. Böttcher\inst{\ref{nwu}}
          \and
          A. Franckowiak\inst{\ref{rub},\ref{rapp}}}

    \institute{Ruhr University Bochum, Faculty of Physics and Astronomy, Astronomical Institute (AIRUB), Universitätsstraße 150, 44801 Bochum, Germany, \email{apel@astro.rub.de}\label{rub}
     \and 
             Centre for Space Research, North-West University, Potchefstroom
            2520, South Africa \label{nwu}
        \and 
             Instituto de Física, Universidade Federal de Goiás, 74690-900, Goiânia, GO, Brazil \label{ufg}
        \and 
             Université Paris Cité, CNRS, Astroparticule et Cosmologie, F-75013, Paris, France \label{apc}
        \and 
             Ruhr Astroparticle and Plasma Physics Center (RAPP Center) \label{rapp}
            }

   %\date{Received XXX; accepted XXX}

% \abstract{}{}{}{}{}
% 5 {} token are mandatory
 
  \abstract
   {Observations of black hole jets closely aligned with our line of sight, known as blazars, have revealed that their X-ray emission can be significantly more polarized than in the radio and optical bands. This chromatic polarization has been suggested to result from an energy-stratified jet. In blazars with X-ray emission dominated by electron synchrotron, the pitch angle between the magnetic field lines and the electrons' velocity vector might also affect the frequency dependence of the polarization degree, but this effect is generally neglected in state-of-the-art models.}
   {In this work, we model the polarization and spectral energy distribution of blazars, accounting for the system's temporal evolution and pitch-angle-dependent electron synchrotron cooling. We investigate the impact of the magnetic field structure and electron distribution on the polarization degree and test whether this can help explain observations of frequency-dependent polarization.} 
   {We extend the numerical simulation code AM$^3$ to include a self-consistent calculation of the polarization of electron synchrotron emission. We simulate various magnetic-field structures and apply these models to the blazar Mrk 421. We compare the predicted flux and polarization with observational data from May 2022. Finally, we present a time-dependent shock-in-jet model to investigate frequency-dependent polarization angle swings.}
   {Our results show how different magnetic-field configurations, including toroidal and random fields, produce distinct energy-dependent polarization signatures. In the case of Mrk 421, the data can be explained by a helical field in the jet. In the shock-in-jet scenario, the X-ray polarization angle changes by $\sim70^\circ$ within 5 h, while remaining almost constant in the optical/infrared.}
   {Our results demonstrate that a self-consistent treatment of the temporal evolution of electrons is essential to model the polarized synchrotron emission from blazars. Pitch-angle-dependent synchrotron cooling can naturally explain chromatic polarization signatures without requiring an energy-stratified jet, allowing instead a co-spatial origin of the emitting particles.}


   \maketitle
   \nolinenumbers

%%%%%%%%%%%%%%%%%%%%%%%%%%%%%%%%%%%%%%%%%%%%%%%%%%%%%%%%%%%%%%
\section{Introduction}
Blazars are a subclass of active galactic nuclei (AGNs) with a relativistic jet close to the observer's line of sight \citep{1995PASP..107..803U}. This leads to strong relativistic Doppler boosting, responsible for the high luminosity and fast variability of blazars. Blazar emission ranges from radio frequencies to gamma rays and originates mainly in non-thermal particle processes. The spectral energy distribution (SED) shows a two-bump structure with a low-energy bump extending from the radio band up to the optical band or, in the case of high-synchrotron-peaked (HSP) blazars, up to X-rays \citep{Abdo_2010}. The high-energy component spans from X-rays up to gamma rays \citep{galaxies7010020}. 

While the low-energy component is generally ascribed to synchrotron emission by non-thermal electrons in the relativistic jet, the origin of the high-energy component remains unclear. Purely leptonic models describe the high-energy emission as a result of synchrotron self-Compton (SSC) or external Compton (EC) process, in which highly relativistic electrons up-scatter their own synchrotron photons or other radiation fields originating outside the jet. In models including non-thermal protons, the high-energy emission can instead be described by proton synchrotron or hadronic processes  \citep{1979ApJ...232...34B, Böttcher_2013}. 

Measurements of multi-wavelength polarization serve as a powerful diagnostic to distinguish between competing emission models for the high-energy radiation \citep{zhang2024revisiting}, and can give us unique insight into the magnetic field structure and the particle acceleration processes \citep[cf. e.g.][]{liodakis2022polarized, galaxies9020027, 2024A&A...681A..12K, di2023discovery}. In this context, we can already gain crucial information from the low-energy emission, since the synchrotron emission from blazars is linearly polarized. 

Observations of HSP blazars usually show an optical polarization degree of a few to tens of percent \citep{galaxies7020046}. While radio and optical polarization have been studied for decades, the launch of the Imaging X-ray Polarimetry Explorer (IXPE) space observatory in 2021 opened a new window for studying multi-wavelength synchrotron polarization from HSP blazars \citep{weisskopf2022imaging}. Observations show that the X-ray polarization degree is typically 10–20\%, which is higher than in the optical band. Also, the polarization angle (or electric vector position angle, EVPA) shows a different behavior in X-rays compared to lower energies. 

For instance, in 2022, IXPE was used to perform an extensive multi-wavelength campaign on the HSP BL Lac object Mrk 421. The analysis shows a strong chromaticity (i.e. energy-dependence) of the polarization degree and a full rotation of the EVPA in X-rays which has not been simultaneously identified in optical/infrared \citep{Di_Gesu_2022, di2023discovery, 2024A&A...684A.127A}. The high observed chromaticity has been explained by an energy-stratified model in which low-energy electrons can propagate farther along the jet into more turbulent regions, while the X-ray-emitting electrons lose their energy more rapidly due to their shorter cooling times, and as a result, they contribute to X-ray emission only close to the acceleration site \citep[e.g.,][]{liodakis2022polarized,abe2024insights,di2023discovery}. This would suggest that the emission in optical/infrared comes from different regions than the emission in X-rays. 

Alternatively, this chromatic behavior might also be explained by the softening of the electron spectrum at higher energies (e.g. \cite{bolis2024multifrequency}). Investigating this requires a self-consistent description of the temporal energy-dependent electron evolution and its impact on the polarization. Synchrotron polarization in blazars has been modeled in several studies \citep[e.g.,][]{zhang2014synchrotron,marscher2014turbulent}. However, geometry affects not only the polarization but also the SED and the energy losses of the electron distribution due to synchrotron cooling. At the same time, most of the public state-of-the-art simulation tools to model the particle evolution in a self-consistent way do not include the geometrical assumptions required to predict polarization signatures. 

In this work, we address this gap by building on the AM$^3$ software \citep{klinger2024am3}, which calculates the temporal evolution of particle populations in astrophysical environments, enabling self-consistent predictions of the multi-wavelength SED and, in the case of hadronic models, neutrinos. We extend AM$^3$ to include synchrotron polarization in a self-consistent manner, including the effect of angle-dependent synchrotron cooling. Because the code is agnostic regarding the geometry of the emission zone, we construct a multi-zone framework to simulate the effect of different magnetic field structures from ordered to random. We investigate how the time-dependent evolution of the electron distribution affects the multi-wavelength polarization, predicting the chromatic behavior of the polarization degree and angle. Finally, we compare our models with observational data from blazar Mrk 421.

This paper is structured as follows: in Section~\ref{theoretical_framework}, we describe the physical background of synchrotron radiation and polarization and our approach to model both. Section~\ref{sec:methods} presents our approach to model complex magnetic field structures, such as toroidal and random magnetic fields, using multi-zone models. We also describe our modeling of the SED and polarization of Mrk 421 and our implementation of the shock-in-jet model. In Section~\ref{results}, we present the results of modeling a random and toroidal magnetic field and their impact on the polarization degree. We show the comparison between the theoretical polarization and SED model with observational data of Mrk 421 during an IXPE measurement in May 2022. Finally, we show the results of a time-dependent simulation of a shock-in-jet model with the best-fit parameters obtained for Mrk 421 in the previous section. In Section~\ref{discussion} we discuss our results and we summarize our work in Section~\ref{summary}.

\section{Theoretical modeling framework}
\label{theoretical_framework}
\subsection{Angle-dependent synchrotron radiation}
\label{sec:syn_rad}

\begin{figure} [htbp!]
    \centering
    \includegraphics[width=0.9\linewidth]{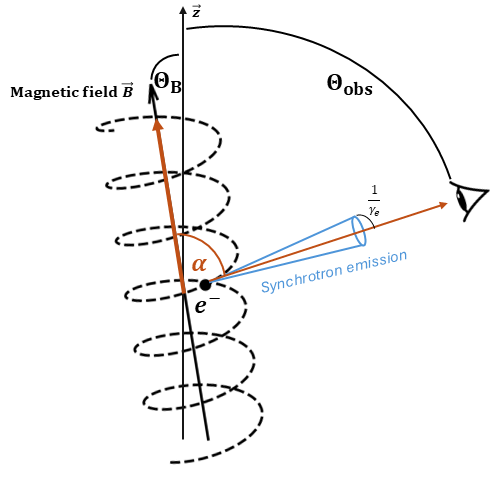}
    \caption{Sketch of the emission geometry. The dashed line shows the electron path around the magnetic field line. The direction of the magnetic field can be parameterized with $\Theta_{\mathrm{B}}$. Throughout this work, the $z$-axis is assumed to be aligned with the jet axis. The electron emits a narrow cone of synchrotron radiation, along its direction of motion, with opening angle $1/\gamma'_{\mathrm{e}}$. Therefore, an observer only receives emission from electrons propagating in the direction of the line of sight, defined by $\Theta_{\mathrm{obs}}$. Following this approach, the pitch angle $\alpha$ can be defined using the line of sight.}
    \label{fig:sketch_geometry}
\end{figure}

The low-energy emission (from radio to soft X-rays) from blazars is associated with synchrotron radiation from relativistic electrons in the jet. A single electron in a magnetic field spirals around the magnetic field lines with a fixed pitch angle $\alpha$, which is the angle between the direction of motion of the electron and the magnetic field line. The electron loses energy in the form of synchrotron radiation, which is emitted in the same direction as the direction of motion of the electron (see  Fig.~\ref{fig:sketch_geometry}). 

Due to relativistic beaming, an observer only sees synchrotron radiation when the line of sight lies inside an emission cone with opening angle $1/\gamma'_\mathrm{e}$, where $\gamma'_\mathrm{e}$\footnote{Parameters with or without prime refer to the values in the jet or observer's frame, respectively.} is the Lorentz factor of the electron. Consequently, radiation from a single electron is observed as a short pulse by an observer at a viewing angle $\theta_{\mathrm{obs}}$ (see Fig.~\ref{fig:sketch_geometry}).
Therefore, in this paper, the observation direction $\hat{\vec{n}} = (\mathrm{sin}(\theta_{\mathrm{obs}}), 0, \mathrm{cos}(\theta_{\mathrm{obs}}))$ and electron propagation direction is considered as equal and the pitch angle can be calculated with:
\begin{equation}
    \label{cos_pitch_angle}
    \mathrm{cos}(\alpha) = \hat{\vec{n}} \cdot \hat{\vec{B}},
\end{equation}
where $\hat{\vec{B}}$ describes the magnetic field direction as follows: 
\begin{equation}
\hat{\vec{B}} =
(\cos\phi_{\mathrm{B}}\sin\theta_{\mathrm{B}},
 \sin\phi_{\mathrm{B}}\sin\theta_{\mathrm{B}},
 \cos\theta_{\mathrm{B}}),
 \label{eq_mag_field}
\end{equation}     
where $\theta_{\mathrm{B}}$ is the angle between the magnetic field and the jet axis and $\phi_{\mathrm{B}}$ is the azimuthal angle.

For a given pitch angle, we have to test first if electrons with slightly different pitch angles contribute to the emission along the line of sight. In Appendix \ref{appendix_pitch_angle}, we show that it is justified to consider only the photons emitted by electrons propagating in the same direction, as the contribution from electrons with different pitch angles is negligible. 

Following this approach, we consider a non-isotropic pitch angle distribution, where the pitch angle is determined by the assumed geometry and viewing angle and is fixed for each individual case. We assume that the electron energy spectrum follows a power law with exponential cutoff $dN/d{\gamma'}_\mathrm{e} \propto {\gamma'}_\mathrm{e}^{-p_\mathrm{e}}\mathrm{exp(-\gamma_\mathrm{e}'/{\gamma'}_\mathrm{e}^\mathrm{cut})}$ with a spectral index denoted as $p_\mathrm{e}$, spanning a range of Lorentz factors above  ${\gamma'}_\mathrm{e}^\mathrm{min}$, while ${\gamma'}_\mathrm{e}^\mathrm{cut}$ defines the characteristic Lorentz factor above which the spectrum decreases exponentially. 

Their energy loss rate depends on their pitch angle $\alpha$ as follows \citep{longair2011high}: 
\begin{equation}
\label{eq_syn_cool}
    -\dot{\gamma'}_\mathrm{e} \propto B'^2 \gamma'^2_\mathrm{e} \mathrm{sin}(\alpha)^2.
\end{equation}
Synchrotron radiation emitted by a population of electrons follows the power spectrum:
\begin{equation}
    \label{syn_power}
    P_{syn}(\nu', \alpha) = \frac{\sqrt{3} e^3 B'~ \mathrm{sin}(\alpha)}{m_e c^2} \int F(x) N_e(\gamma_e')d{\gamma'_e},
\end{equation}
where $x = \nu'/\nu'_c$ with the critical frequency $\nu_c' = \frac{3eB'\gamma_e'^2 \mathrm{sin}(\alpha)}{4 \pi m_e c}$, $B'$ is the magnetic field strength, $N_e$ describes the electron distribution, $F(x) = x\int_x^\infty K_{5/3}(\zeta) d\zeta$, and $K_{5/3}$ is the modified Bessel function of second kind with order 5/3.

\subsection{Synchrotron polarization}
\label{syn_pol}
Synchrotron radiation is linearly polarized \citep{rybicki1979radiative}. The synchrotron emission power can be divided into two components: $P_{\perp}$, perpendicular to the magnetic field projected onto the plane of the sky, and $P_{\parallel}$, the parallel component. The polarization degree of the radiation emitted by the electron distribution can then be calculated as follows:
\begin{equation}
    \label{pol_degree}
    \Pi_{\mathrm{theo}}(\nu') = \frac{P_{\perp}(\nu') - P_{\parallel}(\nu')}{P_{\perp}(\nu') + P_{\parallel}(\nu')} = \frac{\int G(x) N_e(\gamma_e')d{\gamma_e'}}{\int F(x)N_e(\gamma_e')d{\gamma_e'}},
\end{equation}
where the denominator corresponds to the total power, as quantified in Eq.~\eqref{syn_power} (excluding the prefactor) and $G(x) = xK_{\frac{2}{3}}(x)$. For an electron energy spectrum following a simple power law with spectral index $p_\mathrm{e}$ and a uniform magnetic field, integrating Eq.~\eqref{pol_degree} yields the relation
\begin{equation}
    \label{initial_PD}
    \Pi_{\mathrm{theo}} = \frac{p_\mathrm{e}+1}{p_\mathrm{e}+\frac{7}{3}}.
\end{equation}
For now, we only consider the optically thin regime, which is to say we neglect the effect of synchrotron self-absorption (SSA). For a given spectral index, Eq.~\eqref{initial_PD} predicts a single, energy-independent polarization degree (e.g. $\Pi= 69\%$ for a spectral index value of $p_\mathrm{e}$ = 2). However, a full description of the polarization degree must also account for time-dependent cooling effects.

\subsection{Polarization modeling with AM\textsuperscript{3}}
\label{synpol_in_am3}
For this work, we extended the numerical simulation software AM$^3$ to study the effect of the magnetic field structure and the temporal evolution of the electron distribution on the synchrotron spectrum. As described in \citet{klinger2024am3}, AM$^3$ solves a system of coupled integro-differential equations which is derived from each species’ Boltzmann equation. The resulting partial differential equation takes the following form:
\begin{equation}
\label{PDE_AM3}
    \frac{\partial n(E,t)}{\partial t} = -\frac{\partial}{\partial E} \left( \dot{E}(E,t) n(E,t) \right) - \alpha(E,t) n(E,t) + Q(E,t),
\end{equation}
where $n(E,t)$ is the differential number density as a function of energy $E$ and time $t$, and the right-hand side consists of terms describing cooling/advection, escape/sink, and injection/source, respectively. 

Each interaction process (such as synchrotron radiation) can be turned on or off via individual switches. The original version of the AM$^3$ software calculates the synchrotron cooling and emission rate for a homogeneous and isotropic magnetic field, which excludes the treatment of geometric effects. For that case, the pitch-angle dependent synchrotron power $P_{\mathrm{syn}}(\nu', \alpha)$ is integrated over the isotropic pitch-angle probability distribution $p(\alpha) = \frac{1}{2}\int_{0}^{\pi}P_{\mathrm{syn}}(\nu', \alpha)~\mathrm{sin}(\alpha)~d\alpha$ \citep{dermer2009high}. 

For this work, we implemented two new switches for pitch-angle-dependent synchrotron emission and polarization. The pitch angle is introduced as an additional parameter of the emission zone and determined from the assumed geometry using Eq.~\eqref{cos_pitch_angle}. The corresponding synchrotron kernel is then used to calculate the photon source term of Eq.~\eqref{PDE_AM3}. 
When enabled, pitch-angle-dependent synchrotron cooling is treated consistently according to Eq.~\eqref{eq_syn_cool}.

\begin{figure} [htbp!]
    \centering
    \includegraphics[width=1.\linewidth]{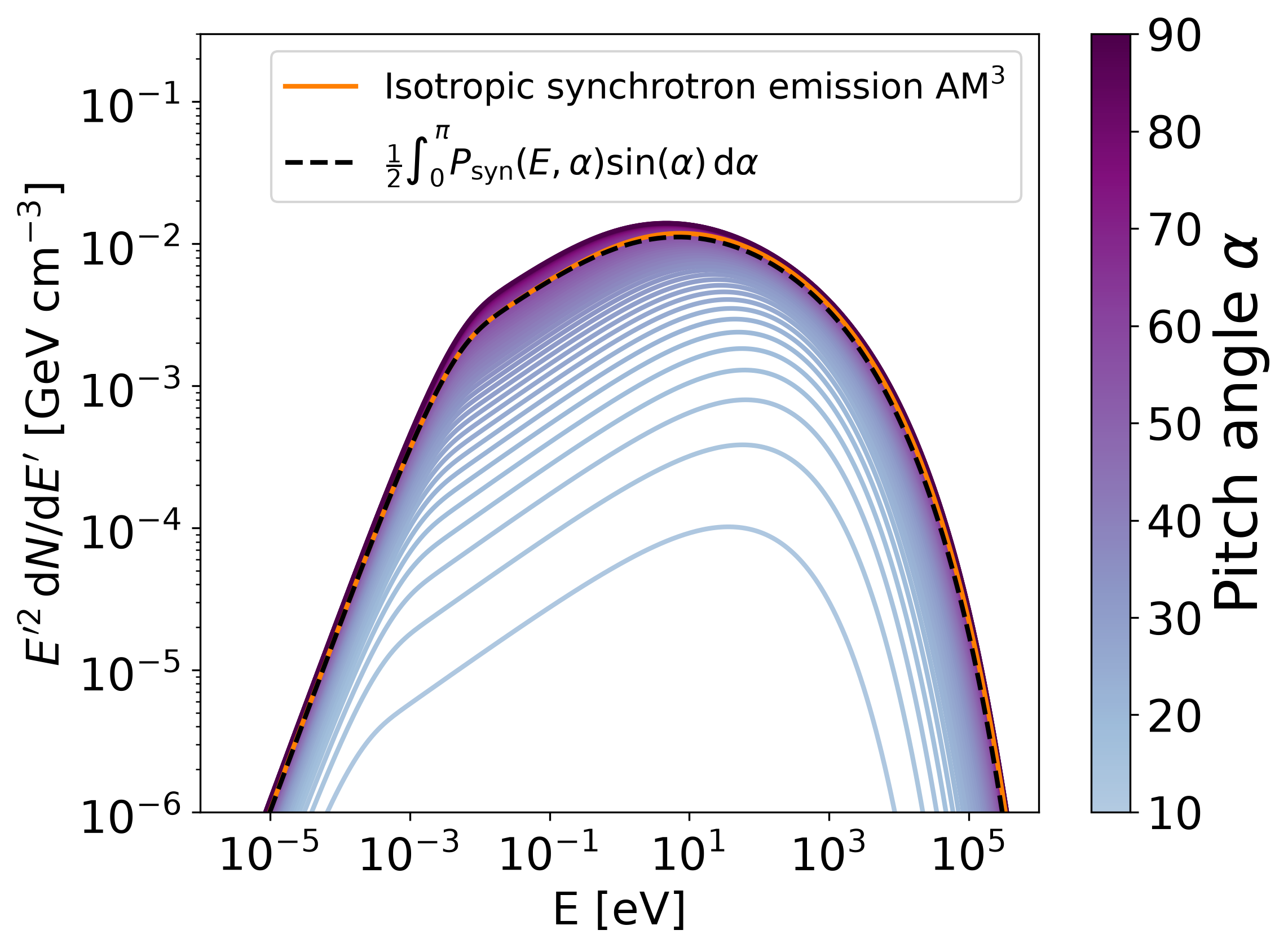}
    \caption{Synchrotron spectra following $P_{\mathrm{syn}}(\nu', \alpha)$ for different pitch angles between 0$^\circ$ and 180$^\circ$. The total emission increases until it reaches a maximum at pitch angle of 90$^\circ$, while the spectral cooling break shifts to lower energies.}
    \label{fig:diff_pitch_angles}
\end{figure}

For effects of comparison between the released AM$^3$ version and the new features, we show in Fig.~\ref{fig:diff_pitch_angles} the synchrotron emission of a steady-state electron distribution for a series of pitch angles, equally spaced between 0$^\circ$ and 180$^\circ$. The leptonic model parameters used in this case are listed in Table \ref{tab:parameter_1}. 

\begin{table*}
    \centering
    \caption{Leptonic model parameters used in Sect.~\ref{synpol_in_am3} and \ref{multi-zone-models}. The bulk Lorentz factor $\Gamma_{\mathrm{b}}$ is not included since the corresponding simulations are performed in the jet frame.}
    \begin{tabular}{lll}
    \toprule
    Parameter & Value & Description \\
    \midrule
    $R^\prime_\mathrm{blob}$, [cm] & $10^{16}$ & Radius of the sperical emission zone region\\
    $B^\prime$, [G] & 0.2 & Strength of the magnetic field\\
    $\gamma_\mathrm{e}^{\prime\mathrm{min}}$ &  $10^{3}$ & Minimal electron Lorentz factor\\
    $\gamma_\mathrm{e}^{\prime\mathrm{cut}}$ & $10^{6}$ & Cutoff electron Lorentz factor\\
    $p_\mathrm{e}$ & 2.3 & Power-law index of the electron energy distribtion\\
    $L^\prime_\mathrm{e}$, [erg s$^{-1}$] & $10^{40}$&  Total electron luminosity\\
    \bottomrule
    \end{tabular}
    
    \label{tab:parameter_1}
\end{table*}

\begin{figure*}[t]
    \centering
    \includegraphics[width=0.87\textwidth]{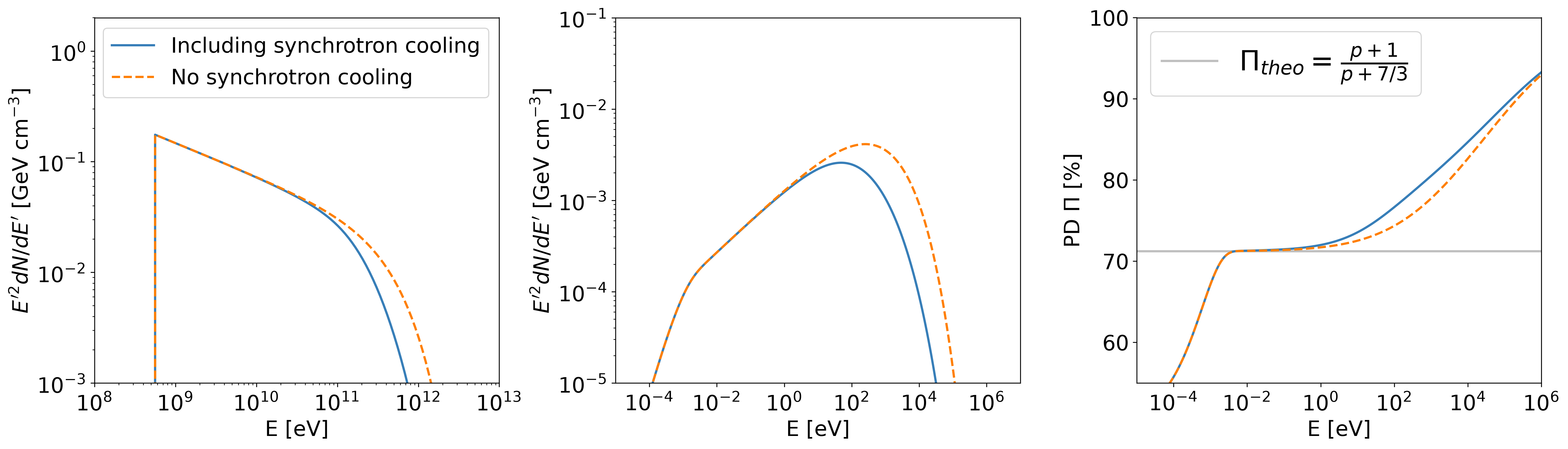}

    \vspace{0.5cm}

    \includegraphics[width=0.87\textwidth]{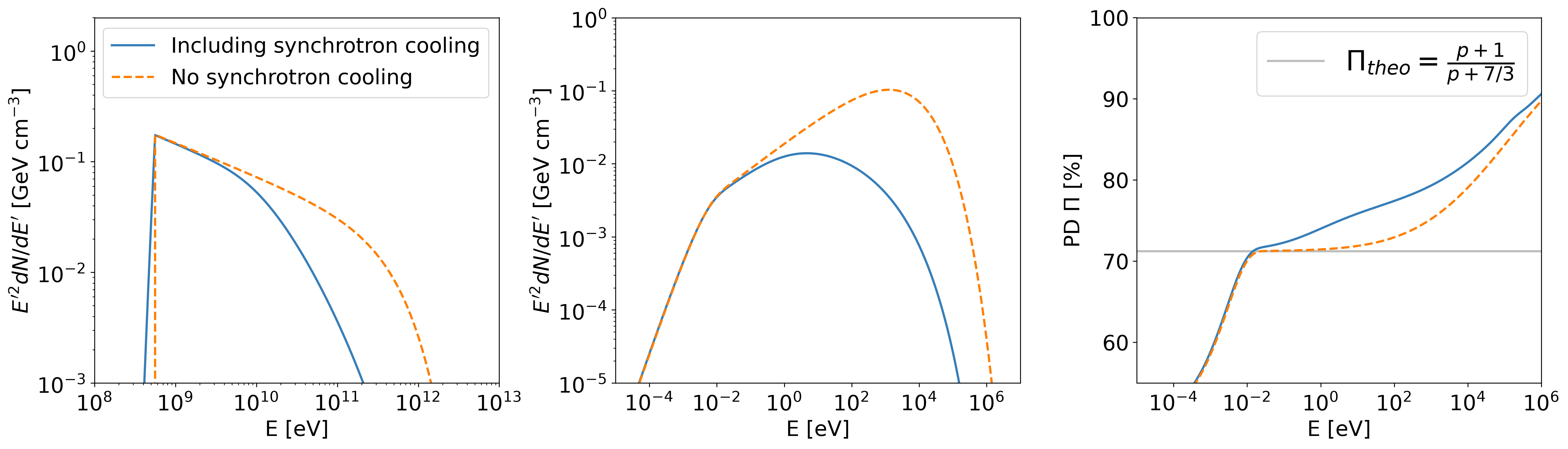}

    \caption{Impact of synchrotron cooling on the polarization of the emission. From left to right: electron energy distribution, synchrotron spectrum, and polarization degree as a function of energy. The solid blue curves represent the case including the energy losses of the electrons due to synchrotron cooling, while the dashed orange curves show the case of an electron distribution without cooling effects. The upper panels show the results for a low pitch angle $\alpha$ = 11$^\circ$, and the lower panels for $\alpha$ = 90$^\circ$. These correspond to magnetic-field directions that are almost parallel and perpendicular to the line of sight, respectively. The parameters describing the electron distribution, the region size, and magnetic field strength are listed in Table \ref{tab:parameter_1}. 
    }
    \label{fig:syn_pol_cooling}
\end{figure*}
Spectra with lower total flux correspond to magnetic-field orientations that are nearly parallel to the line of sight, while the spectra with higher fluxes come from electrons moving more perpendicularly to the field lines. The latter also cool at the fastest rate, which leads to a cooling break at lower energies, as can be seen in Fig.~\ref{fig:diff_pitch_angles}. The individual pitch-angle-dependent synchrotron spectra are integrated over the isotropic pitch-angle distribution (dashed line). The resulting angle-averaged spectrum reproduces the result obtained with the original version of AM$^3$, which assumes an isotropic and homogeneous magnetic field (orange curve).

If the synchrotron polarization process is switched on in addition to the angle-dependent synchrotron emission process, AM$^3$ calculates $G(x)$. The polarization degree can then be obtained by dividing $G(x)$ by $F(x)$ (see Eq.~\eqref{pol_degree}). Previous studies often approximate the polarization degree inside a uniform magnetic field by the theoretical value $\Pi_{\mathrm{theo}}$, which depends solely on the electron spectral index as given in Eq.~\eqref{initial_PD} \citep[e.g.,][]{marscher2014turbulent}. However, electrons undergo continuous energy losses, leading to a time-dependent evolution of their energy spectrum. As demonstrated in Fig.~\ref{fig:syn_pol_cooling}, accounting for this temporal evolution yields significantly different polarization degrees compared to the simplified approach. In Fig.~\ref{fig:syn_pol_cooling}, we show two steady-state electron spectra (left), their respective synchrotron emission spectrum (middle) and the polarization degree of the emission (right),  for the pitch angle of 11$^\circ$ and 90$^\circ$ (upper and lower row), with and without accounting for cooling, using the parameter values listed in Table \ref{tab:parameter_1}. Without cooling effects (dashed orange curves), the polarization degree is constant up to the cutoff energy, with $\Pi$ around 71\% for our chosen spectral index $p_{\mathrm{e}}$ = 2.3. Above the cutoff energy, the exponential cutoff steepens the spectrum dramatically, leading to a fast increase in the polarization degree up to 100\% at the highest energies, which follows directly from Eq.~\eqref{initial_PD}. However, because this extreme polarization is only realized at energies above the cutoff, it does not significantly contribute to the observed emission. On the contrary, when accounting for cooling (solid blue curves), the electron spectrum steepens above a certain break energy, which can also be seen in the emitted synchrotron spectrum. As we can see by comparing the two right panels in Fig.~\ref{fig:syn_pol_cooling}, this energy-dependent increase in the polarization degree has an onset at lower energies for larger pitch angles, owing to the stronger effect of cooling. Even in the case of a small pitch angle $\alpha=11^\circ$, this effect  is significant at energies below the emission peak (upper-right panel).

For large angles, this effect leads to values that are around 10\% higher than the expected values according to Eq.~\eqref{initial_PD}. These examples show that the temporal evolution of electrons alters the polarization significantly, and even for small pitch angles, chromaticity effects from angle-dependent synchrotron cooling must be taken into account.

\section{Methodology}
\label{sec:methods}

\subsection{Constructing complex magnetic field structures}
\label{multi-zone-models}
So far, we considered only simplified scenarios with a uniform magnetic field direction; we now introduce our methodology to include more complex magnetic field configurations. As we discussed in Sect.~\ref{syn_pol}, synchrotron emission from an electron population with $p_\mathrm{e}=2.3$ radiating in a uniform magnetic field would be expected to have a polarization degree of $\sim71$\%, assuming an optically thin source. Observations of blazars show polarization degrees significantly lower than that \citep[e.g.,][]{galaxies7020046}. This difference can be due to several reasons, including depolarization effects from turbulent fields or symmetry effects that lead to decrease of polarization. To investigate more complex magnetic field structures, we use multi-zone models and set a fixed observation angle and a different magnetic field direction in each zone. 

The Stokes parameters for each individual zone can be written as
\begin{equation}
\label{stokes}
    (I_i, \, Q_i, \, U_i) = L_{\nu,i} \cdot (1, \, \Pi_i\mathrm{cos}(2 \chi_i), \, \Pi_i\mathrm{sin}(2 \chi_i)),
\end{equation}
where the index $i$ denotes the corresponding emission zone, $ L_{\nu,i}$ is the luminosity, $\Pi_i$ is the polarization degree calculated with Eq.~\eqref{pol_degree} and $\chi_i$ is the electric vector position angle (EVPA). Since the electric field vector is always perpendicular to the projected magnetic field onto the plane of the sky, the EVPA can be determined for each zone. Using the Stokes parameters, the total polarization degree is: 
\begin{equation}
\label{eq_PD_tot}
\Pi_{\mathrm{tot}} = \frac{\sqrt{(\sum_i^N Q_i)^2 + (\sum_i^N U_i)^2}}{\sum_i^N I_i},
\end{equation}
and the total EVPA can be calculated with:
\begin{equation}
\label{eq_tot_EVPA}
    \chi_{\mathrm{tot}} = \frac{1}{2}\arctan\frac{ \sum_i^N U_i}{\sum_i^N Q_i},
\end{equation}
where $N$ is the number of zones. Using these equations, we can analyze the effects of complex magnetic field topologies such as a toroidal or a random magnetic field.

For a toroidal field, we set $\theta_{\mathrm{B}}=90^\circ$, corresponding to a magnetic field perpendicular to the jet axis, Eq.~\eqref{eq_mag_field} reduces to the toroidal magnetic field configuration:
\begin{equation}
\hat{\vec{B}}= (\mathrm{cos}(\phi_\mathrm{B,i}), \mathrm{sin}(\phi_\mathrm{B,i}), 0),    
\end{equation}
where $\phi_\mathrm{B,i}$ represents the azimuthal angle in each zone. Such a symmetric magnetic-field structure does not require a large number of zones, as the result obtained with N=20 is consistent with that for N=100. We therefore adopt N=20 zones, with $\phi_\mathrm{B,i}$ equally spaced between 0 and $2\pi$. Testing different viewing angles, the results are presented in Section \ref{toroidal_results}.

While a toroidal magnetic field represents a highly ordered field configuration, turbulent magnetic-field components are also expected to be present in blazar jets \citep[e.g.][]{marscher2014turbulent,Guo:2017lxn,galaxies9020027}, which can lead to depolarization. This may explain why blazars often exhibit relatively low polarization degrees, despite synchrotron radiation being intrinsically highly polarized. The depolarization effect can be quantified as follows. Let us consider $N$ zones, each with a randomly generated $\theta_{\mathrm{B}}$ and $\phi_{\mathrm{B}}$ using Eq.~\eqref{eq_mag_field}. The expected value of the total observed polarization can be quantified by \citep{zhang2024revisiting}:
\begin{equation}
\label{eq_depolarization_factor}
    \Pi_{\mathrm{tot}} = \frac{\Pi_{\mathrm{theo}}}{\sqrt{N}},
\end{equation}
where $\Pi_{\mathrm{theo}}$ is the theoretical maximum of the polarization degree, given by Eq.~\eqref{initial_PD}. We investigate this depolarization effect by simulating random magnetic-field configurations with different numbers of zones. The results are presented in Section \ref{random_results}.

\subsection{Helical magnetic field model for Mrk 421}
\label{helical_field_methods}
Multiple studies have suggested the presence of a helical magnetic field in the jet of Mrk~421 \citep{di2023discovery}, multiple other blazars \citep[][]{Gabuzda:2004kc,zhang2014synchrotron,2024A&A...681A..12K}, and radio galaxies \citep{EventHorizonTelescope:2021iqj,Nikonov:2023gih}. Here, we focus on Mrk 421 and we simulate the effect of such a helical structure on the energy-dependent polarization of the emission, using our multi-zone approach described in the previous section. In 2022, \citet{2024A&A...684A.127A} reported on a broadband observational campaign of Mrk 421, covering the radio-to-gamma-ray range, including the first IXPE polarization measurements in X-rays. The observations in May revealed a low flux state at high energies, a typical behavior in optical, UV and MeV-GeV gamma-rays, and moderate daily variability in X-rays. The polarization degree in X-rays, $\Pi = 15 \pm 2\%$, was found to be higher than in the optical/IR band, $\Pi\approx3\%$. 

To investigate this chromatic behavior of Mrk 421, we have tested whether the SED and the polarization degree observed in May 2022 can be explained by a distribution of electrons located in a single region with a helical magnetic field. For the SED modeling, we constrain the initial guess of the model parameters to typical value ranges, based on previous studies of Mrk 421 \citep[e.g.][]{zhang2014synchrotron, abe2026time, 2024A&A...684A.127A}, assuming an SSC origin of the gamma-ray data. Since modeling the SED for a range of values of $\theta_{\mathrm{B}}$ would be computationally expensive, and our primary aim was to obtain approximate leptonic parameters for the subsequent polarization modeling, we adopted the isotropic electron pitch-angle distribution implemented in the original AM$^3$ code for the SED modeling. This approximation is also justified by the fact that averaging over the different pitch angles in a helical magnetic field configuration results in a similar total synchrotron flux as assuming an isotropic electron pitch-angle distribution. We optimized the model by fitting to data via reduced chi-squared minimization, using a genetic algorithm \citep[cf. e.g.][]{kramer2017genetic} to search the parameter space of the model within the established value ranges following the procedure in \citet{Rodrigues2019Blazars}. We added systematic uncertainties to the statistical uncertainties of the optical -- X-ray data. This was particularly important for the X-ray data, where the statistical uncertainties were very small and would otherwise have dominated the $\chi^2_{\mathrm{red.}}$ minimization, causing the fit to be overly constrained by the X-ray data.

The best-fit parameters were then used for the polarization modeling. Since our line of sight is nearly along the jet, we assumed the typical observation angle $\theta_{\mathrm{obs}} \sim 1/\Gamma_{\mathrm{b}}$. Due to relativistic aberration, the viewing angle transforms to  $\theta'_{\mathrm{obs}}$ = 90$^\circ$ in the comoving frame. Similar to the toroidal field, Eq.~\eqref{eq_mag_field} was used to define the helical field, with the azimuthal angle $\phi_{\mathrm{B}, i}$ equally distributed across the 20 zones. To find the angle $\theta_{\mathrm{B}}$, we plotted the polarization degree for helical fields with 10 equally spaced $\theta_{\mathrm{B}}$ values between 0$^\circ$ and 90$^\circ$ and compared the results with the observational data. The results are shown in Section \ref{Mkr421_data}.

\subsection{Simulating a time-dependent shock}
\label{shock_in_jet_methods}
\begin{figure} [htbp!]
    \centering
    \includegraphics[width=1.\linewidth]{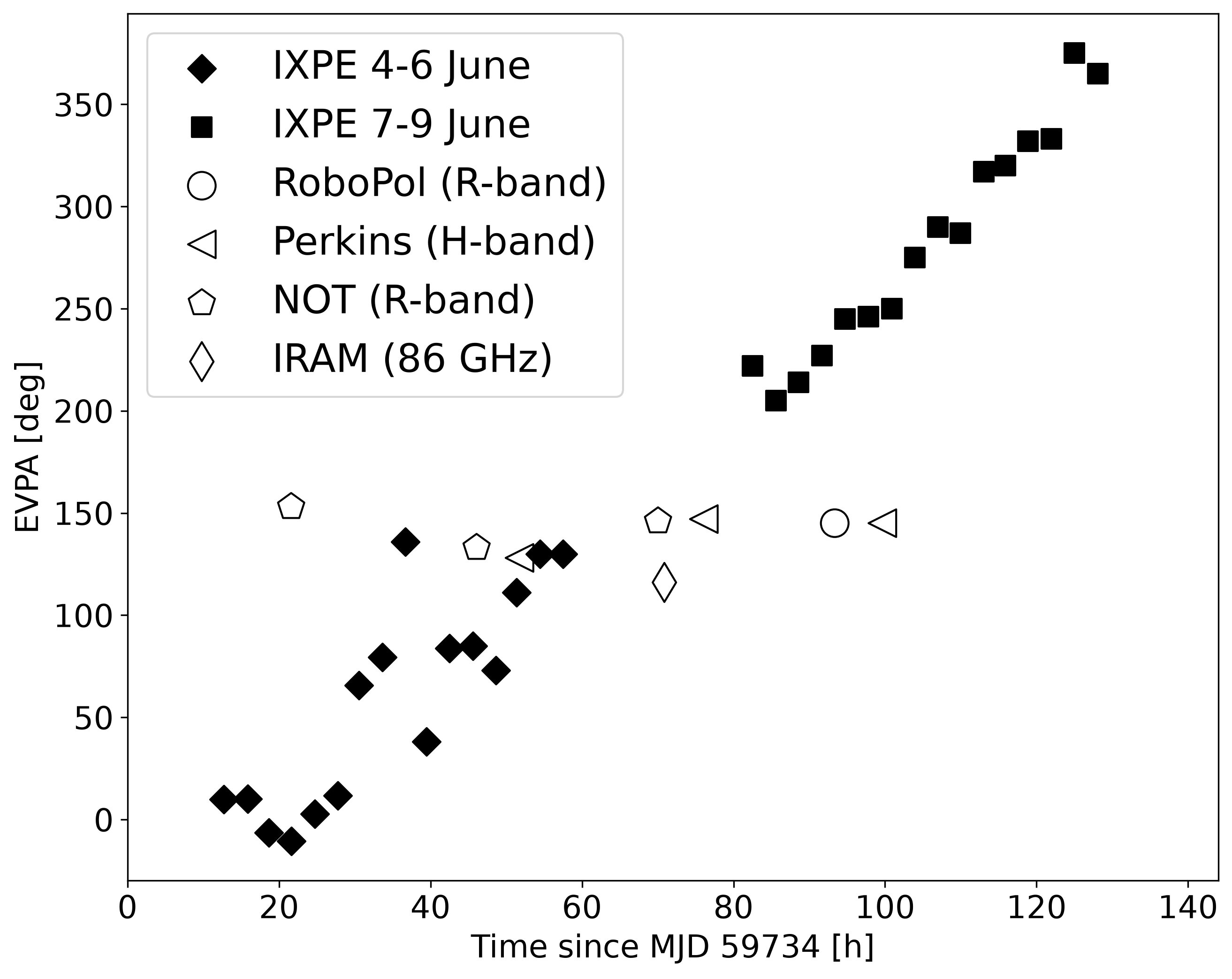}
    \caption{Observations of Mrk 421 in June 2022: EVPA evolution over time at different wavelengths. Data taken from \citet{2024A&A...684A.127A}.
    }
    \label{fig:PA_data}
\end{figure}

In June 2022, IXPE observed Mrk 421 and detected an EVPA rotation ($\geq$ 360$^\circ$) in X-rays accompanied by an increase in the X-ray flux. In contrast, the optical/IR data does not reveal a significant change in EVPA over time. The observations were divided into two epochs, June 4–6 and June 7–9. The corresponding data is shown in Fig.~\ref{fig:PA_data}. The observations in June 2022 show a polarization degree in X-rays of 10\% while the polarization degree in optical/IR remained at a similar level ($\Pi \sim$ 5\%)  as in May 2022. \citet{di2023discovery} propose that the X-rays are produced in a compact inner spine region, whereas the optical emission originates from a surrounding sheath. In that scenario, the observed X-ray EVPA rotation can be explained by a shock propagating along the jet's helical magnetic field, enhancing the X-ray emission by injecting high-energy particles. 

Instead of invoking two separate emission regions (the compact inner spine and the outer sheath), as suggested by the authors, we investigated whether a single magnetic-field structure can lead to different EVPA behavior at X-ray and lower energies. We study the temporal EVPA evolution driven by changes in the electron distribution, with the aim of testing whether this scenario can reproduce the observed differences in EVPA behavior across energies. Reproducing the complete EVPA rotation ($\geq 360^\circ$) of Mrk~421 is beyond the scope of this study.

\begin{figure} [htbp!]
    \centering
    \includegraphics[width=1.\linewidth]{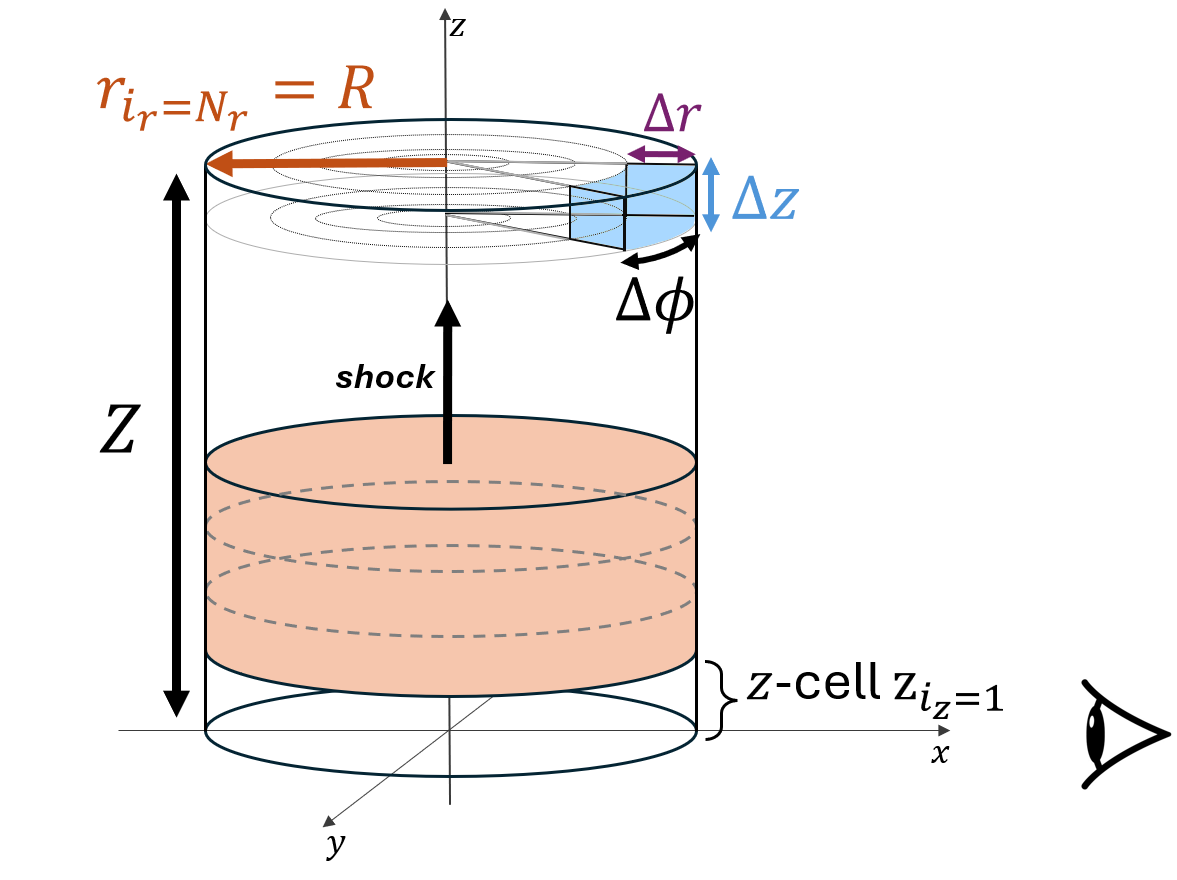}
    \caption{Geometry of the emission region adapted from \cite{zhang2014synchrotron}. The cylindrical emission region has a length of $Z = 10^{16}$ cm. The total radius $R$ results from the parameter search described in Sect. \ref{helical_field_methods}. The cylinder is divided into smaller zones, one of which is highlighted in blue, with dimensions $\Delta r = R/27$, $\Delta z = Z/30$ and $\Delta \phi = 2\pi/120$. The shock is a cylinder with radius $R$ and length $Z/10 = 3\cdot \Delta z$ moving along the $z$-axis with the speed of light. The observer is located along the direction $\hat{\vec{n}} = (1, 0, 0)$.}
    \label{fig:sketch_shockinjet}
\end{figure}

Inspired by \citet{zhang2014synchrotron}, we tested a scenario in which a shock propagates through an emission region pervaded by a helical magnetic field, continuously injecting additional high-energy electrons. Following their procedure, we assumed a cylindrical geometry for the entire emission region, as illustrated in Fig.~\ref{fig:sketch_shockinjet}. The total length of the emission region is fixed to $Z = 10^{16}$ cm. We adopted the radius determined with the methods described in Sect.~\ref{helical_field_methods}. We discretized the cylinder into $N_r$ = 27 radial cells, $N_z$ = 30 cells along the $z$-axis, which is aligned with the jet axis, and $N_\phi$ = 120 cells along the azimuthal direction. The shock is a cylindrical region as well, with the same radius as the whole emission region and length $Z/10 = 10^{15}$~cm, propagating along the axis at the speed of light $c$. Since the shock length spans three of the 30 cells along the jet axis (see Fig.~\ref{fig:sketch_shockinjet}), the duration of the enhanced particle injection is given by $\Delta t_{\mathrm{shock}} = 3 \cdot Z/30/c$ $\sim$ 9 h. 


To mimic the passage of a strong shock, which locally increases the particle injection rate and accelerates electrons up to higher energies, we used the following method: we first evolved the electron distribution down to the steady state, using the geometry and leptonic model parameters found with the methods described in Section \ref{helical_field_methods}. Once the system reached steady state, we increased the electron injection luminosity by a factor of 10, and hardened the electron spectrum by reducing the spectral index to $p_{\mathrm{e}}$ = 1. This allowed us to simulate a flare while remaining within a physically plausible parameter range. However, the shock parameters were chosen to demonstrate the effect of enhanced particle injection and are not intended to provide a fit to a specific observation. The newly injected electrons evolve with time, following the AM$^3$ time-dependent solver formalism \citep[cf.][]{klinger2024am3}, and after $\Delta t_{\mathrm{shock}}$, the injection parameters are reset to their quiescent values. This emulates the fact that the shock propagates downstream, leaving the region.

Since we assume the same conditions in each $z$-cell, the simulation with AM$^3$ is done for one $z$-cell, divided by 120 $\phi$-cells, each with the corresponding $\phi_{\mathrm{B}, i_\phi}$ from Eq.~\eqref{eq_mag_field}. After the simulation, we account for the propagation of the shock front along $z$ and the light travel time effects (LTTE), which introduce time delays in the observed emission due to the different locations of the zones within the emission region. In our geometry, this introduces two independent delay components. First, we calculate the Stokes parameters $Q(t')$, $U(t')$, and $I(t')$ using Eq.~\eqref{stokes}, where $t'$ denotes the local comoving time. Since the shock propagates along the jet axis ($z$-axis) with the speed of light $v_{\mathrm{shock}} = c$, the time shift depends on the position along the $z$-axis. Consequently, the temporal evolution of the Stokes parameters in each plane along the $z$-axis is obtained by shifting the simulation time according to:
\begin{equation}
t^{\mathrm{delay}}_{i_z} = t' + \frac{z_{i_z}}{v_{\mathrm{shock}}},
\end{equation}
with $i_z \in [1, N_z]$. Following the standard formalism \citep[e.g.,][]{rybicki1979radiative}, the observed arrival time is determined by the comoving time $t'$ of the cell and the projection of its position vector $\hat{\vec{r}}$ onto the line of sight $\hat{\vec{n}}$, such that $t_{\mathrm{obs}} = t' - \frac{\hat{\vec{n}} \cdot \hat{\vec{r}}}{c}$, with $\hat{\vec{r}} = (r_{i_r}\cdot \mathrm{cos(\phi_{i_\phi}}), r_{i_r}\cdot \mathrm{sin(\phi_{i_\phi}}), z_{i_z})$ and $\hat{\vec{n}} = (1,0,0)$. To avoid negative time indices, we shift the reference point of the delay to the front edge of the cylinder at $\vec{r_0} = (R, 0,0)$. The time delay is given by:
\begin{equation}
    t_{i_r, i_\phi}^{\mathrm{delay}} = \frac{R}{c} -\frac{r_{i_r}}{c}\cos\left(\phi_{i_\phi}\right),
\end{equation}
where $i_r \in [1, N_r]$ and $i_\phi \in [1, N_\phi]$ denote the radial and azimuthal cell indices, respectively, $r_{i_r}$ is the radial distance from the cylinder axis, $\phi_{i_\phi}$ is the azimuthal angle, and $c$ is the speed of light. 

Since the volume element in cylindrical coordinates scales as $dV \propto r~dr~d\phi~dz$, cells at larger radial distances represent larger volumes and are therefore weighted accordingly. We therefore multiply the Stokes parameters of each radial cell by the normalized weight
\begin{equation}
    w_{i_r} = \frac{r_{i_r}}{\sum_{i_r = 1}^{N_{r}} r_{i_r}}.
\end{equation}

The time delays are applied to the Stokes parameters of each cell, so we obtain $Q(t_{\mathrm{obs}}), U(t_{\mathrm{obs}})$ and $I(t_{\mathrm{obs}})$ with $t_{\mathrm{obs}} = t' + t^{\mathrm{delay}}_{i_z} + t_{i_r, i_\phi}^{\mathrm{delay}}$. The delayed Stokes parameters are then summed over all cells to obtain the total Stokes parameters, from which the polarization degree and EVPA are calculated using Eqs.~\eqref{eq_PD_tot} and \eqref{eq_tot_EVPA}. The results are shown in Sect.~\ref{EVPA_shockinjet_results}.

\section{Results}
\label{results}

\subsection{Signatures of complex magnetic field structures}
\label{Effect_mag_field_results}

We first apply the magnetic-field configurations and methods described in Sect.~\ref{multi-zone-models} to investigate their impact on the synchrotron SED and polarization signatures.

\subsubsection{Toroidal field}
\label{toroidal_results}

\begin{figure}[htbp!]
    \centering
    \includegraphics[width=.9\linewidth]{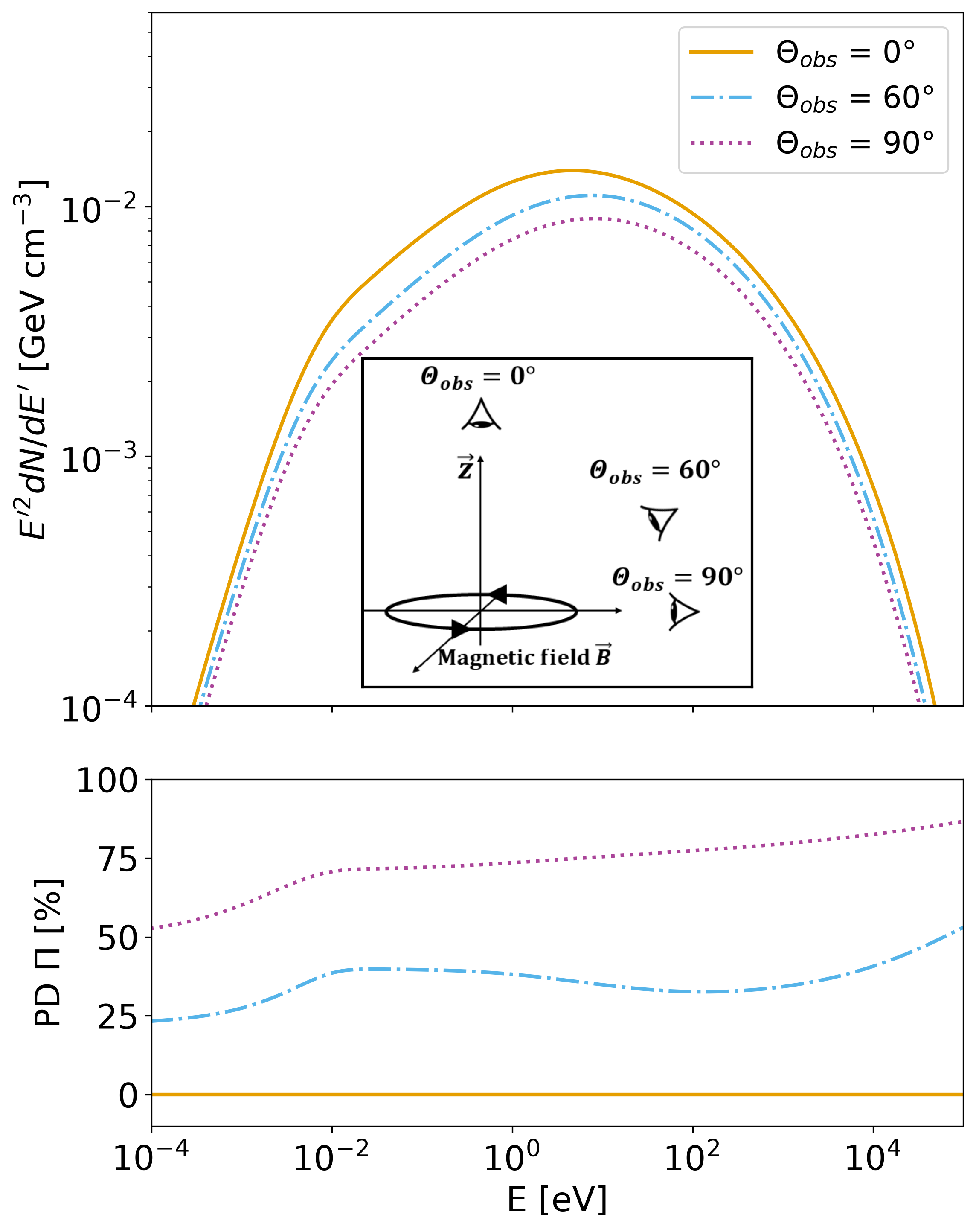}
    \caption{Synchrotron emission and polarization for different viewing angles of a toroidal magnetic field. The upper panel shows the synchrotron spectra, while the lower panel shows the corresponding polarization degree as a function of photon energy.}
    \label{fig:toroidal}
\end{figure}

The toroidal field is an axisymmetric configuration that can either strongly increase or reduce the net polarization degree, depending on the viewing angle. To portray the effect of a toroidal magnetic field configuration on the observed emission, let us consider three observers at different observation angles: $\theta_\mathrm{obs}=0^\circ$, $60^\circ$, and $90^\circ$, as sketched in the upper panel of Fig.~\ref{fig:toroidal}. Let us also consider the parameters from Table \ref{tab:parameter_1}.

We show the corresponding observed synchrotron spectra and polarization curves in Fig.~\ref{fig:toroidal}. As we know from the previous discussion, the observed emission comes mainly from electrons moving toward the observer, while the contribution from electrons moving in other directions can be neglected. The observed flux is proportional to the magnitude of the perpendicular component of the magnetic field, $B_\perp$. The case of $\theta_\mathrm{obs}=0^\circ$ maximizes this component, yielding the highest observed synchrotron flux (solid orange curve in the upper panel). At the same time, because at every point the projected field $\vec B_{\perp,i}$ has the same magnitude and the electrons have the same pitch angle, all zones contribute equally to the emission. The EVPA in each zone is always perpendicular to $\vec{B}_{\perp, i}$. Due to the symmetry of the magnetic field, the opposite EVPAs from these regions cause the corresponding Stokes parameters (defined in Eq.~\eqref{stokes}) to cancel out pairwise, so the net polarization vanishes (solid orange curve in the lower panel).

Instead, by observing the system with $\theta_\mathrm{obs}=60^\circ$, we probe electrons at a range of pitch angles lower than $90^\circ$, decreasing the total observed flux and increasing the polarization degree (blue, dash-dotted curves in Fig.~\ref{fig:toroidal}. The polarization degree also displays a non-trivial behavior as a function of energy: it reaches 40\% around $10^{-2}$~eV, then it decreases to about 30\% at $10^{2}$~eV, and then increases again at energies beyond the spectral cutoff. This non-monotonic behavior would not be expected from the simple treatment of Eq.~\eqref{initial_PD}, and it emerges from the fact that at $\theta_\mathrm{obs}=60^\circ$, the emission originates in zones with different pitch angles, each aligned differently with the line of sight. 

Finally, $\theta_\mathrm{obs} = 90^\circ$ represents an observer lying along the toroidal plane, which breaks the symmetry maximally. This viewing angle minimizes the perpendicular component of the toroidal magnetic field relative to the line of sight, yielding the minimum possible observed synchrotron flux (dotted purple curve in the upper panel of Fig.~\ref{fig:toroidal}. It also maximizes the polarization degree, as those zones where the field is perpendicular to the line of sight (which contribute highly to the emission) have similar EVPA, while all the other zones with a different EVPA contribute minimally, since the magnetic field is more parallel to the line of sight. Only in this case does the polarization degree correspond to the uniform magnetic field scenario of Eq.~\eqref{initial_PD}, in this case $\Pi \approx71\%$.

\subsubsection{Random magnetic field}
\label{random_results}

\begin{figure}[htbp!]
    \centering
    \includegraphics[width=.9\linewidth]{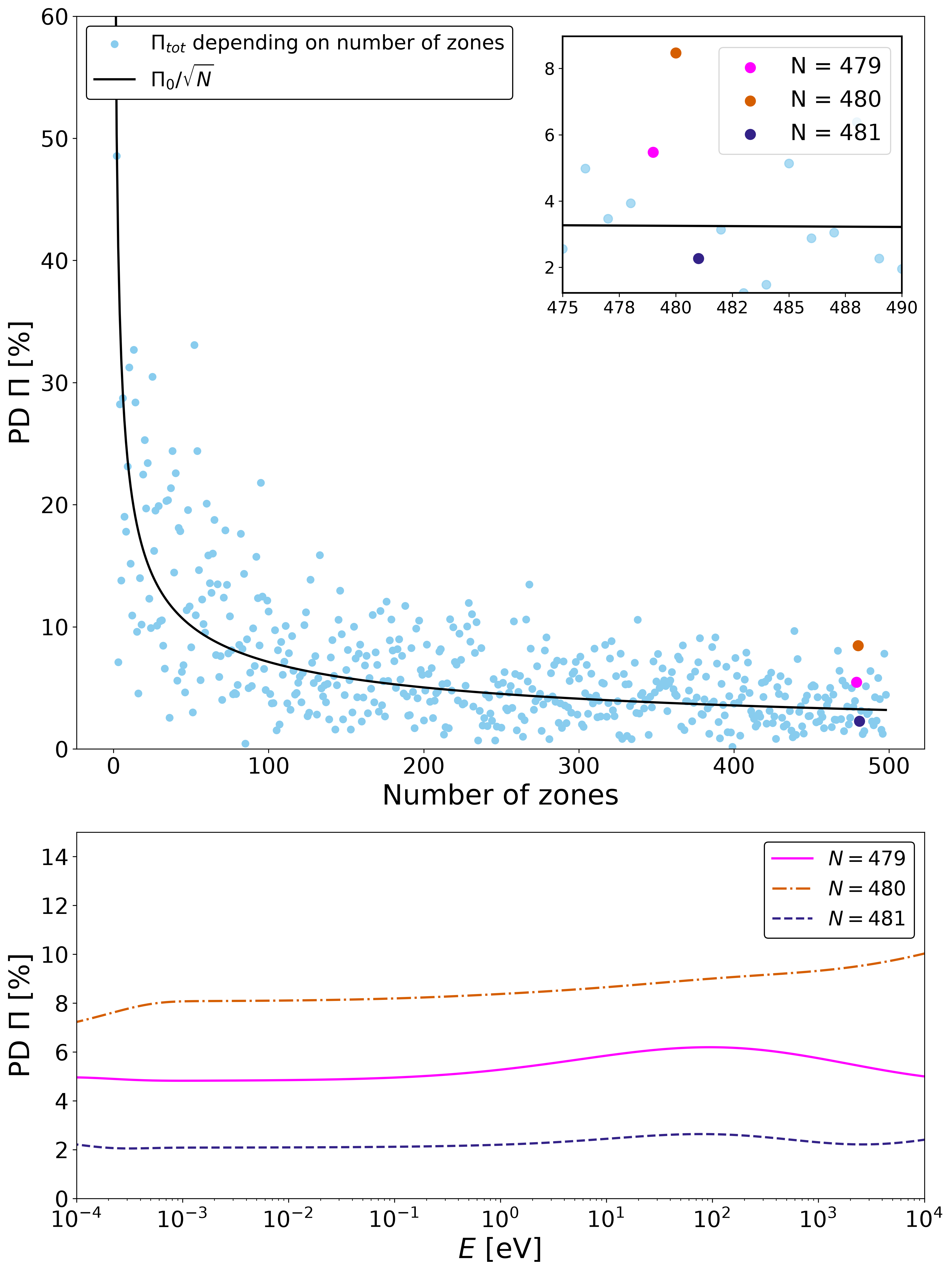}
    \caption{Upper panel: Net polarization degree as a function of the number of zones for a random magnetic-field configuration. The solid line shows the expected theoretical depolarization factor. Three cases with a similar number of zones but significantly different polarization degrees are highlighted. Lower panel: Polarization degree as a function of photon energy for the three highlighted cases.}
    \label{fig:chaotic}
\end{figure}

Based on the consideration in Sect. \ref{multi-zone-models}, the depolarization effect of a random magnetic field is illustrated in the upper panel of Fig.~\ref{fig:chaotic}. It shows the total polarization degree of synchrotron radiation with energy 2.3~eV (optical), depending on the number of zones. Each point represents an independent simulation of a random magnetic field with the corresponding number of zones, with the simulation restarted and a new random field generated for each number of zones. The points generally follow the theoretically predicted depolarization behavior (solid line in the upper panel); however, even for a large number of zones, there is a non-negligible spread. To have a closer look at the polarization degree, we picked three results with a high number of zones, namely 479, 480, and 481 (highlighted points). The lower panel of Fig.~\ref{fig:chaotic} shows the corresponding polarization degree as a function of photon energy. The theoretically expected polarization degree according to Eq.~\eqref{eq_depolarization_factor} is $\Pi_{\mathrm{tot}} \sim$ 3\%. While the polarization degree for $N=481$ remains around 2\%, the case with $N=480$ reaches values of around 9\%. For $N=479$ (pink curve), the polarization degree starts around 5\% and slightly increases around $10^2$ eV. We see that differences in the polarization degree between different energies are still present. However, the random magnetic field configuration appears to reduce the chromaticity, resulting in less pronounced differences in the polarization degree across energies. The maximum difference in the polarization degree across the different energies is around 4\% (see dash-dotted curve).

\subsection{Modeling the steady-state polarization of Mrk 421}
\label{Mkr421_data}

\begin{figure} [htbp!]
    \centering
    \includegraphics[width=1.\linewidth]{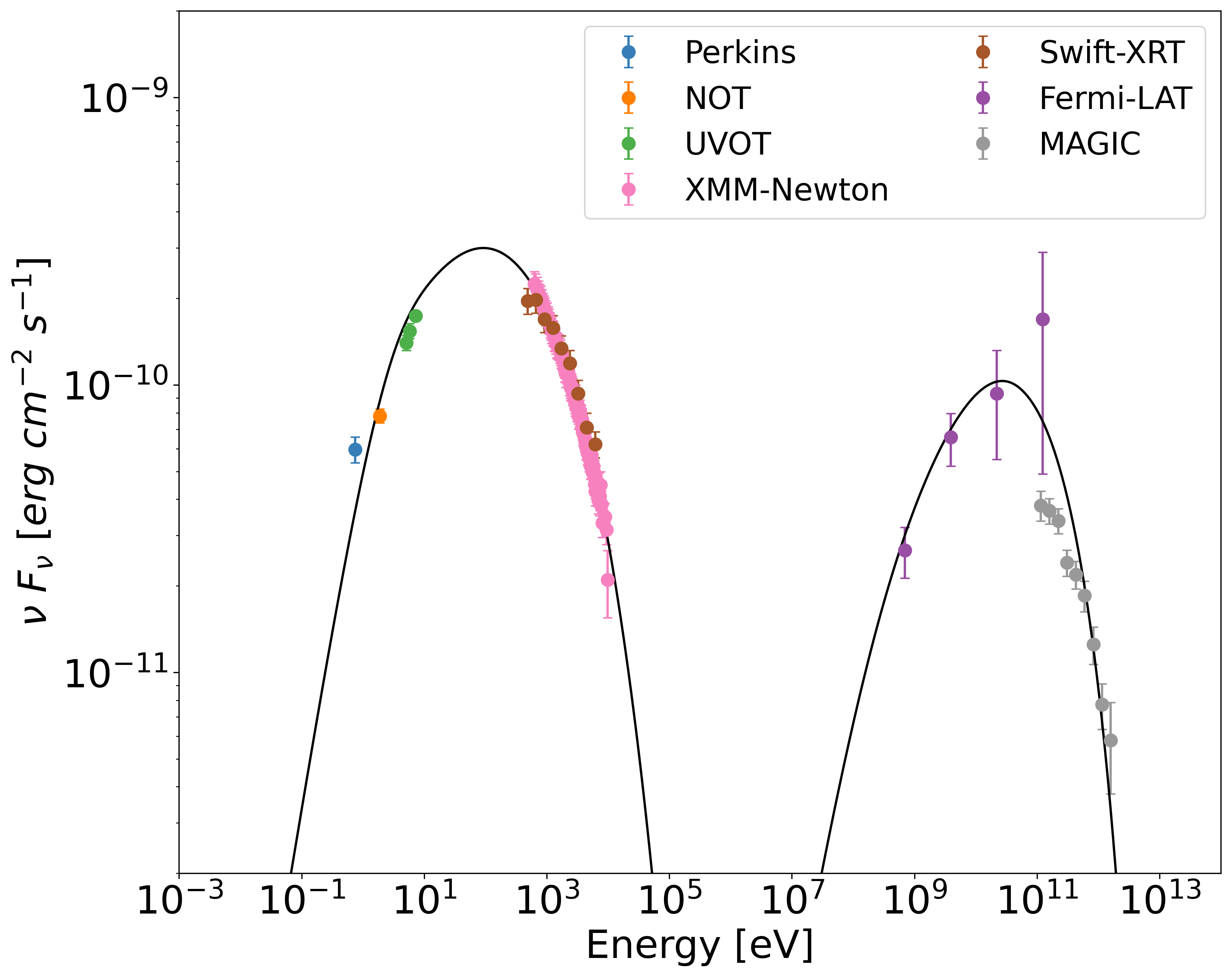}
    \caption{Broadband multi-wavelength flux measurements from Mrk 421 and model prediction (black curve). The data correspond to the  observation period from May 4-6, 2022. The Fermi-LAT data (0.3 - 300 GeV) was averaged over a 7-day period centered around the IXPE observation time. The data is taken from \citet{2024A&A...684A.127A}.}
    \label{fig:sed_fit}
\end{figure}

Following the methodology described in Sect. \ref{helical_field_methods}, the SED and polarization degree of Mrk 421 were modeled. We list the best-fit parameters in Table \ref{tab:lep_pars}, and we show the resulting SED in Fig.~\ref{fig:sed_fit}. The model provides a good description of the X-ray spectrum and the available gamma-ray flux measurements within their respective uncertainties with $\chi^2_{\mathrm{red.}}$ = 2.1 (including the systematics of the instruments as descibed in Sect. \ref{helical_field_methods}).

\begin{figure} [htbp!]
    \centering
    \includegraphics[width=1.\linewidth]{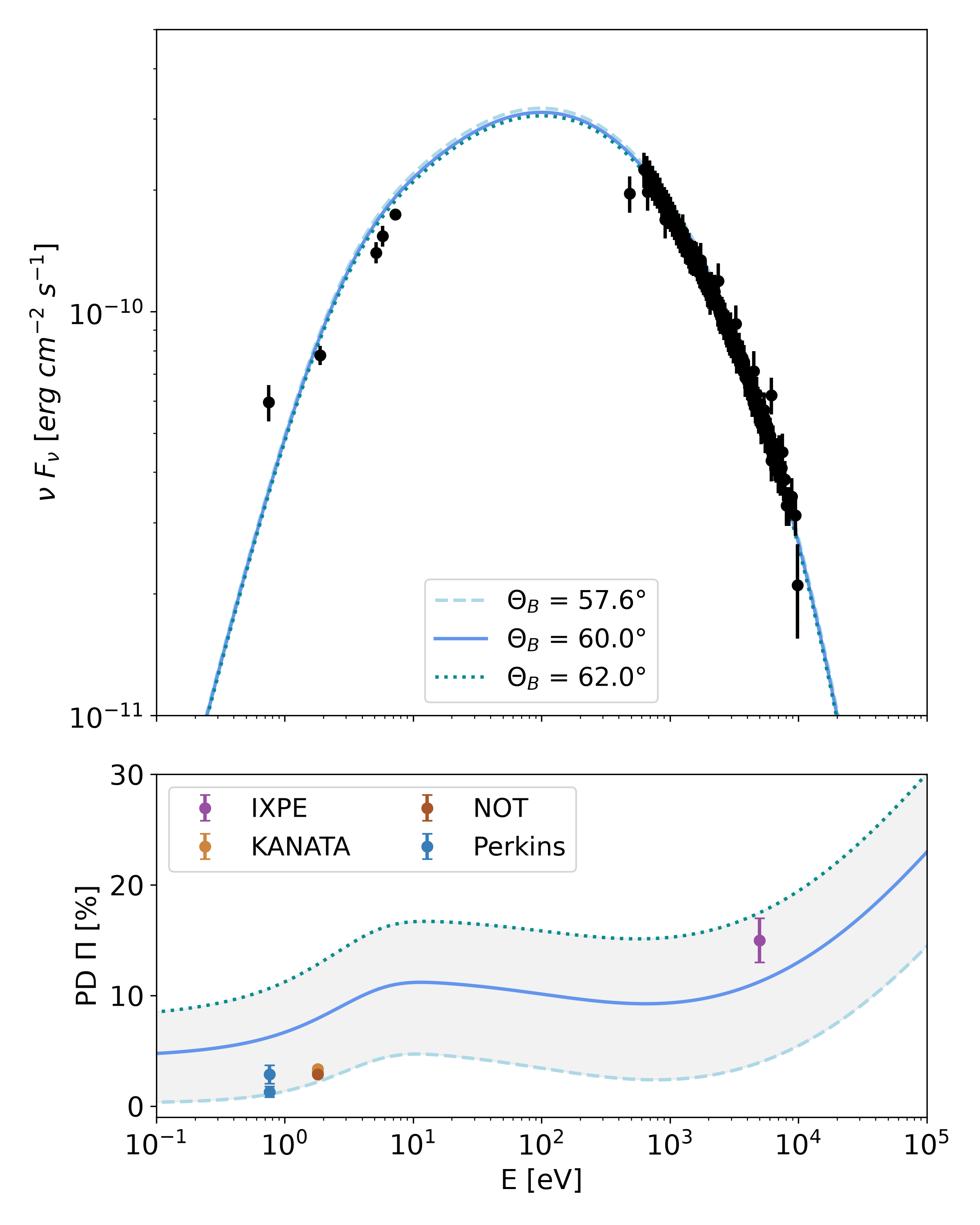}
    \caption{Synchrotron SED (upper panel) and polarization degree (lower panel) for Mrk 421, including data points from May 4-6, 2022 and model predictions for helical magnetic field configurations with varying $\theta_{\mathrm{B}}$, as indicated in the legend.}
    \label{fig:pol_data_fit}
\end{figure}

Assuming a helical magnetic field, we used the same leptonic model parameters for modeling the synchrotron polarization of Mrk 421 as those used to describe the SED. Accounting for relativistic aberration, we fixed the viewing angle to $\theta'_{\mathrm{obs}} \sim 90$$^\circ$ since we assume $\theta_{\mathrm{B}} \sim 1/\Gamma_{\rm b}$. For $\theta'_{\mathrm{B}} \sim 60$$^\circ$, the model prediction shows the best agreement with the polarization data. 
The results in the lower panel of Fig.~\ref{fig:pol_data_fit} show the polarization degree as function of energy for $\theta_{\mathrm{B}} = 60$$^\circ$ surrounded by a shaded region indicating results of 57.6$^\circ$ and 62$^\circ$. The polarization data lies within this region of $\theta_{\mathrm{B}} \sim$ 60$^\circ$ $\pm$ 2$^\circ$. The polarization degree predicted by the model naturally increases with energy, as synchrotron cooling and the exponential cutoff modify the electron distribution at high energies. 

\begin{table} [htbp!]
    \centering
    \caption{Best-fit parameters obtained for the leptonic model fit to the SED and polarization degree of Mrk 421 in the quiescent state.}
    \begin{tabular}{ll}
    \toprule
    Parameter & Value\\
    \midrule
    $R^\prime_\mathrm{blob}$, [cm] &$6.93 \times 10^{15}$\\
    $B^\prime$, [G] & 0.19\\
    $\Gamma_\mathrm{b}$ &21.50\\
    $\gamma_\mathrm{e}^{\prime\mathrm{min}}$ & $5.89 \times 10^{3}$\\
    $\gamma_\mathrm{e}^{\prime\mathrm{cut}}$ & $1.46 \times 10^{5}$\\
    $p_\mathrm{e}$ & 2.10\\
    $L^\prime_\mathrm{e}$, [erg s$^{-1}$] &  $3.10 \times 10^{40}$\\
    
    \midrule
    $\chi^2_{\mathrm{red.}}$ &  2.1
    \end{tabular}
    
    \label{tab:lep_pars}
\end{table}

\subsection{Time-dependent shock evolution}
\label{EVPA_shockinjet_results}

Finally, we apply the time-dependent shock scenario described in Sect.~\ref{shock_in_jet_methods}, in which a harder and brighter electron population is injected as the shock propagates through the helical magnetic field. We investigate the resulting temporal evolution of the polarization across different photon energies. 
\begin{figure}[htbp!]

\centering
    \includegraphics[width=1.\linewidth]{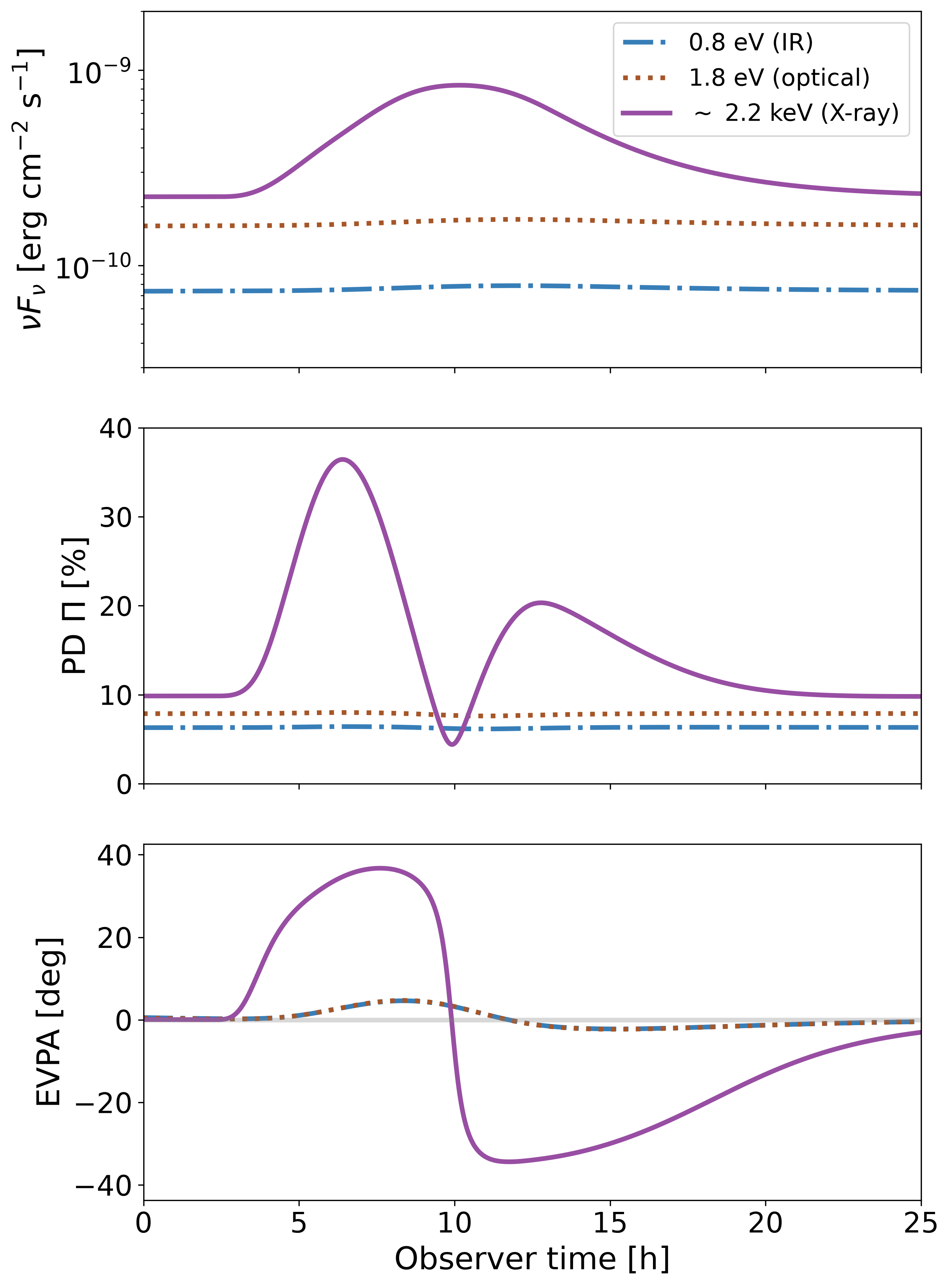}
     
        \caption{Results of the shock-in-jet model. From top to bottom: light curves, polarization degree, and EVPA evolution in the IR, optical, and X-ray bands. The light curves show the temporal evolution of the emitted flux, while the polarization signatures illustrate the corresponding evolution of the polarization degree and EVPA induced by the propagating shock. The shock is represented by enhanced electron injection propagating through the emission region.}
        \label{fig:shock_in_jet}
\end{figure}

\begin{figure*} [htbp!]
    \centering
    \includegraphics[width=1.\linewidth]{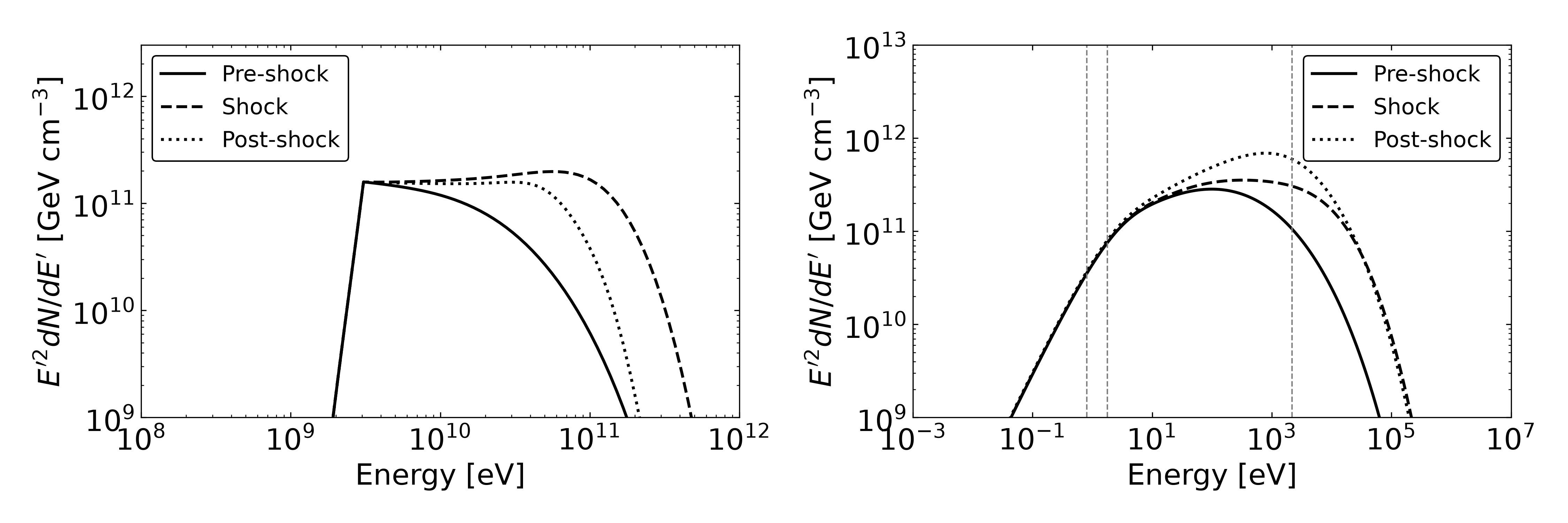}
    \caption{Evolution of the electron energy distribution (left) and the corresponding synchrotron spectrum (right) in the emission frame  before, during, and after the shock passage. The vertical dashed lines in the right plot indicate the same photon energies shown in Fig.~\ref{fig:shock_in_jet}: 0.8, 1.8, and $\sim$ 2.2 keV.}
    \label{fig:syn_and_elec_shock}
\end{figure*}

The upper panel of Fig.~\ref{fig:shock_in_jet} shows the light curves in the IR, optical, and X-ray bands. The middle and lower panels present the polarization signatures, with the middle panel showing the polarization degree as a function of time and the lower panel showing the corresponding EVPA evolution. For all plots, we transformed the time to the observer's frame. If the EVPA is $\chi_{\mathrm{tot}}$ = 0$^\circ$, the electric field vector is parallel to the projection of the jet axis onto the plane of the sky in this model. Increasing the EVPA corresponds to a counter-clockwise rotation of the electric field vector. With $\theta_{\mathrm{B}}$ = 60$^\circ$, the toroidal component dominates over the poloidal component of the helical magnetic field. The ratio between both components is given by $\tan(\theta_{\mathrm{B}})=B_{\mathrm{tor}}/B_{\mathrm{pol}}$, resulting in $B_{\mathrm{tor}}/B_{\mathrm{pol}}\approx1.7$. Therefore, the EVPA in Fig.~\ref{fig:shock_in_jet} starts with 0$^\circ$, corresponding to an electric field vector aligned with the jet axis. Due to LTTE, the observer first receives photons from the near side of the emission region. This observed emission already contains a contribution from the shock-affected electron distribution. Since the EVPA is perpendicular to the projected magnetic field, the observed EVPA initially rotates counter-clockwise. As the shock propagates further downstream, emission from more distant regions with different projected magnetic-field orientations reaches the observer, causing the EVPA to reverse its direction and rotate clockwise. In total, the EVPA changes from +35$^\circ$ to -35$^\circ$, corresponding to a total rotation of 70$^\circ$.

At the same time, this asymmetric configuration leads to a high peak in polarization degree with $\Pi \sim$ 35\%. After about 9 h, the polarization degree briefly drops while the EVPA reverses its direction and begins to rotate clockwise. This marks the transition phase during which the emission from the near and far sides of the emission region contributes almost equally, partially decreasing the net polarization. As the shock propagates further downstream, the emission becomes dominated by the more distant zones, causing the polarization degree to increase again. In both the polarization degree and the EVPA, we obtain an asymmetric behavior i.e. the polarization degree of the second bump is lower. This is because the newly injected electrons still evolve in the closer zones, even after the shock has already passed through them.

At lower energies, the EVPA varies slightly in the beginning. This is because $\gamma_\mathrm{e}^{\prime\mathrm{min}}$ and $\gamma_\mathrm{e}^{\prime\mathrm{cut}}$ remain unchanged during the shock passage. Therefore, additional electrons are injected across the entire energy range, including lower-energy electrons. Although their contribution to the total synchrotron emission is small, the enhanced emission from these electrons slightly changes the relative weighting of zones with different magnetic-field orientations, resulting in a minor EVPA variation. However, the polarization degree remains unaffected, since the additional contribution from the shock-accelerated low-energy electrons is negligible compared to the emission from the non-shock regions. These regions contain the majority of the low-energy electrons and, therefore, dominate the synchrotron emission at lower energies.

In Fig.~\ref{fig:syn_and_elec_shock}, we show the resulting electron and photon spectra before, during, and after the shock in the emission frame. The post-shock spectrum corresponds to a time of $\sim$ 23~h after the shock and illustrates the subsequent evolution of the electron population. 
The synchrotron peak evolves throughout the shock scenario: during the shock phase, the increased injection of high-energy electrons shifts the peak towards higher energies, enhancing the X-ray emission. During the post-shock phase, synchrotron cooling causes the peak to shift towards lower energies.

\section{Discussion}
\label{discussion}
In this section, we discuss the results of Sect.~\ref{results}. By modeling polarization, we particularly investigate the  strong energy dependence of the polarization degree and EVPA. The observed differences between the optical and X-ray polarization properties have often been interpreted as evidence for an energy-stratified jet, in which the high-energy-emitting particles remain confined to compact regions close to the acceleration site due to their short cooling times while low-energy-emitting particles propagate to more turbulent regions.

However, in this work, we modeled the time- and energy-dependent synchrotron polarization in a self-consistent manner. While pitch-angle dependent synchrotron emission has already been considered in previous studies \citep[e.g.][]{galaxies4040045}, we extended this approach by including the pitch-angle dependence of synchrotron cooling. As the results for toroidal and turbulent magnetic field configurations in Sect.~\ref{Effect_mag_field_results} show, this effect has a significant impact on the difference between low- and high-energy polarization while all emission is assumed to originate from the same electron distribution. 
The results for the toroidal field show how the viewing angle can dramatically change the observed polarization degree in a structured-magnetic-field scenario, and how it can introduce significant chromaticity. For instance, in the last scenario with $\theta_\mathrm{obs} = 90^\circ$, the polarization degree changes from about 70\% at eV (optical) to about 80\% at 10~keV (X-rays). For the random magnetic field configuration, we showed the impact of the number of zones on the total polarization degree. This example illustrates that even a turbulent magnetic field sampled over many zones can yield a large polarization degree. At the same time, the chromaticity is weaker than in the case of an ordered magnetic field.

In Sect.~\ref{Mkr421_data}, we modeled the SED of Mrk 421 from May 4-6 2022 with AM$^3$. This has already been done before in \cite{2024A&A...684A.127A}, where the X-ray emission and the very-high-energy gamma-ray emission detected by MAGIC were interpreted as originating from a single SSC-emitting region. However, their one-zone model was not intended to reproduce the entire SED. Instead, based on the energy stratification suggested by the polarization data, they assumed that the optical/UV and high-energy gamma ray emission observed by Fermi-LAT receive significant contributions from larger, spatially separated regions of the jet and therefore purposely underestimated these components. In contrast, our approach aims to describe the full broadband emission by connecting both spectral components within a single co-spatial framework. Additionally, while \citet{2024A&A...684A.127A} used a broken power law for the electron energy spectrum, AM$^3$ calculates the cooling break in a self-consistent way by considering the temporal electron evolution. Comparing the best-fit parameters, we found a higher magnetic field strength with $B$ = 0.19\,G compared to $B$ = 0.042 G found by \citet{2024A&A...684A.127A}. Furthermore, the bulk Lorentz factor of $\Gamma_{\mathrm{b}}$ = 21.5 is smaller in our case compared to the $\Gamma_{\mathrm{b}}$ = 30 used in their work.

Using the same parameters obtained from the SED modeling (see Table \ref{tab:lep_pars}), we modeled the polarization degree for a helical magnetic field and compared it with the observational data. We assumed that all polarization signatures came from synchrotron radiation of an electron distribution. From a small parameter scan, we found the best agreement between model and data with $\Theta_{\mathrm{B}}$ = 60$^\circ$, which represents a helical field with a slightly more dominant toroidal component. 

While our model overestimates the low-energy polarization degree by 3-5\% and underestimates the IXPE polarization degree by approximately 3\%, the data lie between the model predictions for $\Theta_{\mathrm{B}}$ = 57.6$^\circ$ and $\Theta_{\mathrm{B}}$ = 62$^\circ$, corresponding to deviations of about $\pm$ 2$^\circ$ from 60$^\circ$. Since a fixed magnetic field angle represents a highly idealized scenario, slight variations in the magnetic field orientation are expected. Additionally, the polarization degree of blazars is often variable on short time scales, which can lead to deviations from the assumed value. Therefore, our model describes well the geometry and processes during the observation period. 

Our model naturally predicts a large difference between the polarization degree in optical/IR compared to the one in X-rays. Around $E$ = 0.8 eV the polarization degree reaches $\Pi$ = 6\% while the value is doubled in X-rays ($E \sim 7$~keV) with $\Pi$ = 12\%. Therefore, the results show that the chromaticity of the polarization degree can be naturally explained by this scenario without splitting the origin of the emitting particles into two zones.

However, it is important to note that the viewing angle has a significant impact on the polarization degree and only slight changes of 1-2$^\circ$ of the viewing angle in the observer's frame can lead to very different outcomes due to relativistic aberration. Therefore, also the bulk Lorentz factor determines the viewing angle in the jet frame. For instance a viewing angle $\Theta_{\mathrm{obs}}$ = 1$^\circ$ corresponds to $\Theta'_{\mathrm{obs}}$ = 55.26$^\circ$ while $\Theta_{\mathrm{obs}}$ = 2$^\circ$ transforms to $\Theta'_{\mathrm{obs}}$ = 92.63$^\circ$ if the same bulk Lorentz factor is assumed for both cases. Therefore, a degeneracy exists in constraining the exact geometry, as different combinations of the viewing angle and the magnetic-field angle with respect to the jet axis can produce similar polarization signatures. For simplification, we used the typical assumption $\Theta_{\mathrm{obs}} \sim 1/\Gamma_{\mathrm{b}}$ in this work. In a more extended study which is beyond the scope of this paper, VLBI measurements of the viewing angle and the orientation of the jet of Mrk 421 could be used to address this issue. With this additional information, one could also compare the resulting EVPA with the data from May 2022 which was not considered in this work, since the focus was on the analysis of the energy dependence of the polarization degree. Additionally, the parameter space of the leptonic model is also highly degenerate, and there are several possible solutions for parameters that describe the SED data equally well \citep{apel2025impact}. Consequently, variations in the leptonic parameters can lead to different predictions for the polarization degree and EVPA. Combining all available time- and energy-dependent photon flux and polarization measurements can provide a clearer understanding of the leptonic parameters that characterize the source \citep{lucchini2019breaking}. While the energy-dependent behavior of the polarization degree only represents a snapshot in time, it would be interesting to investigate whether this configuration can also simultaneously explain the observed temporal variability of the polarization degree, or whether an additional turbulent component is required.

In Sect.~\ref{EVPA_shockinjet_results} we studied the time- and energy-dependent behavior of the EVPA. The example of the data set from June 2022 illustrates the typical case where the temporal evolution of the EVPA differs significantly across energy bands. The data showed an EVPA rotation in X-rays while the low-energy data stayed constant. In \citet{di2023discovery}, the authors interpret the rotation in June 2022 as a result of a shock propagating along a helical magnetic field. In their scenario, the X-ray emission originates from the inner spine region, where the helical field is present, while the optical/radio emission is produced in the surrounding regions of the jet. Our results show that significant changes in the X-ray EVPA together with a constant EVPA at lower energies can also arise from a single magnetic-field structure, without requiring an energy-stratified model with separate emission regions. This process led to an angle rotation of $\sim$ 70$^\circ$, while the low-energy EVPA varied only slightly. 

This shock-in-jet scenario, in which the injection of high-energy electrons is increased, has been already simulated by \citet{zhang2014synchrotron}. They presented a detailed analysis of time- and energy-dependent synchrotron polarization signatures connected with synchrotron and high-energy flares in Mrk 421 and PKS 1510-089. Overall, we found a similar behavior of the polarization signatures for this specific scenario. Our results for the polarization degree as a function of time also show a two-bump structure with a higher peak in the beginning. Their results also show that the EVPA in low energies exhibits only slight or no variations while the EVPA in X-rays carries out a larger swing from 60$^\circ$ to 120$^\circ$. The differences are mainly caused by the choice of geometry and leptonic model parameters. 

Overall, this scenario is able to describe an EVPA behavior that differs in energy and simultaneously, explains a flare in UV - X-rays. However, the polarization degree reaches relatively high values of around 35\% and 20\%. While such high polarization degrees have already been detected in blazars (e.g. 30\% from PKS 2155–304 \citep{kouch2024ixpe}), the X-ray polarization degree measured by IXPE for Mrk 421 during the EVPA rotation was only $\Pi$ = 10\%. If we average the PD resulting from the shock-in-jet model  over a time interval of around 25 h, the PD is $\Pi$ = 15\%, as contributions with  different EVPAs partially cancel in the Stokes parameters. This highlights the crucial role of the time resolution when comparing model predictions with polarization observations.

Since several factors can influence the temporal evolution of the EVPA, a future extension of this model could investigate whether complete EVPA rotations can be reproduced within a self-consistent framework. Furthermore, simultaneous SED observations during the rotation period could be modeled to constrain the temporal evolution of the physical parameters and provide additional constraints for the polarization modeling. It is also important to note that relativistic effects on the EVPA were not taken into account, although \citet{lyutikov2003polarization} showed that such effects can lead to EVPA rotation. In \citet{peirson2018polarization}, it was shown that these effects also enhance the differences between low- and high-energy polarization signatures. Therefore, future studies could analyze how this alters the results for the time-dependent polarization signatures.

\section{Summary and conclusions}
\label{summary}

In this work, we investigated the impact of particle evolution and magnetic field geometry in blazars on the synchrotron SED and polarization signatures using the numerical simulation code AM$^3$. In particular, we extended the existing framework from \citet{klinger2024am3} by implementing a method to model polarization and by including pitch-angle-dependent synchrotron emission and cooling, allowing us to consistently model the temporal evolution of the electron distribution together with the resulting polarization signatures.

We simulated idealized magnetic field configurations, including uniform, toroidal, and turbulent magnetic fields, and modeled both the synchrotron SED and the polarization degree. The results for a toroidal magnetic field show how the polarization degree depends on the viewing angle. For turbulent magnetic fields, increasing the number of zones, decreases the polarization degree, approaching the case of a turbulent isotropic magnetic field. While a low polarization degree is often associated with turbulent fields, our examples demonstrate how the net polarization can also be reduced by geometrical effects.

Assuming a helical magnetic field, we applied our model to observations of Mrk 421 from May 2022. Using the same best-fit parameters obtained from the SED modeling, we predicted the polarization degree and compared it with the observations. The best agreement was found for a helical magnetic field with an angle of $\theta_{B}$ = 60$^\circ$ relative to the jet axis. The model describes the data well by allowing for small variations in the magnetic field structure of approximately $\theta_{B}= 60^\circ \pm 2^\circ$.

To investigate the origin of chromatic EVPA variations, we modeled a shock propagating through a helical magnetic field by continuously injecting a more energetic electron distribution along the shock region. In this scenario, the self-consistent description of the electrons evolving after the shock and light travel time effects naturally produces an EVPA rotation of about 70$^\circ$, while the optical/IR EVPA remains nearly constant. 

Our results demonstrate that a strong energy dependence of both the polarization degree and the EVPA can arise naturally with a single emission region containing one coherent magnetic field structure. In contrast to the commonly adopted interpretation of an energy-stratified jet with spatially separated emission regions and different magnetic field structures, our results show that the combined effects of pitch-angle-dependent synchrotron cooling, different magnetic field geometries and light travel time effects can explain the observed chromatic polarization signatures. Therefore, a self-consistent treatment of polarization is essential for obtaining a deeper understanding of the underlying physical processes in blazar jets. 

In future work, we will extend this framework to include polarization calculations at higher energies. This will provide an additional tool for distinguishing between leptonic and hadronic emission scenarios. In the end, this extension of AM$^3$ will be made publicly available after a forth-coming dedicated technical paper and can be applied not only to blazars but also to other astrophysical sources exhibiting polarized emission.

\newpage

\begin{acknowledgements}
We thank Cosimo Nigro for valuable discussions and ideas regarding the SED and polarization modeling of Mrk 421. F.A. acknowledges support by the Helmholtz Weizmann Research School on Multimessenger Astronomy. A.F. and F.A. acknowledge the support from the DFG via the Collaborative Research Center SFB1491 \textit{Cosmic Interacting Matters - From Source to Signal}.
V.B.M. gratefully acknowledges the National Council for Scientific and Technological Development (CNPq) for the Research Productivity Fellowship (process no. 312015/2026-7).
X.R. acknowledges support by the investment program ‘France 2030’ launched by the French Government and implemented by the University Paris Cité as part of its program ‘Initiative d’excellence’ IdEx (ANR-18-IDEX-0001), which also funded the HERMES: multi-messengers of the Earth and the Universe project that contributed to this work. 
This paper is supported by the European Union’s Horizon Europe research and innovation programme under grant agreement No 101131928, project ACME. This research has made use of the VizieR catalogue access tool, CDS, Strasbourg, France (DOI : 10.26093/cds/vizier). The original description of the VizieR service was published in 2000, A\&AS 143, 23.
\end{acknowledgements}

\bibliographystyle{bibtex/aa} % style aa.bst
\bibliography{biblio}

\begin{appendix}

\section{Distribution of pitch angles}
\label{appendix_pitch_angle}

In Sect.~\ref{sec:syn_rad}, it was discussed that synchrotron emission is emitted within a cone around the electron velocity direction. Therefore, it is necessary to test whether the assumption that the observer only receives photons from electrons with one specific pitch angle is justified. The opening angle of the cone depends on the electron energy with $\Theta \sim 1/\gamma'_\mathrm{e}$. In the following, it will be investigated if photons from emission cones of electrons with different pitch angles have to be taken into account. For this, we have to consider the synchrotron radiation of the following distribution of pitch angles:
\begin{equation}
    P'^{\mathrm{avg}}_{\mathrm{syn}}(x) = \int_{\alpha-1/\gamma'_{\mathrm{e}}}^{\alpha+1/\gamma'_{\mathrm{e}}} P'_{\mathrm{syn}}(x,\alpha) d\alpha,
\end{equation}
where $\alpha$ is the fixed pitch angle depending on the LOS [cf. Eq.~\eqref{cos_pitch_angle}] and $x$ is the same as in Eq.~\eqref{syn_power}. For the synchrotron power, we use $P'_{\mathrm{syn}}(x,\alpha) \approx B'\mathrm{sin}(\alpha)\cdot F(x)$ for the regarding $\gamma'_{\mathrm{e}}$ that is used for the opening cone instead of integrating over a power law distribution.  Since the opening cone becomes wider for lower electron energies, the largest deviation from the fixed pitch-angle approximation is expected for low-energy electrons. We consider electrons with $\gamma'_{\mathrm{e}} \in [10, 30, 50, 100]$. The corresponding opening angles are $\Theta \in$ [5.7$^\circ$, 1.9$^\circ$, 1.1$^\circ$, 0.6$^\circ$]. For the fixed pitch angle $\alpha_0$, we choose a small angle of 6$^\circ$, as this maximizes the relative difference of $\mathrm{sin}(\alpha)$ within the pitch angle range.

\begin{figure} [htbp!]
    \centering
    \includegraphics[width=1.\linewidth]{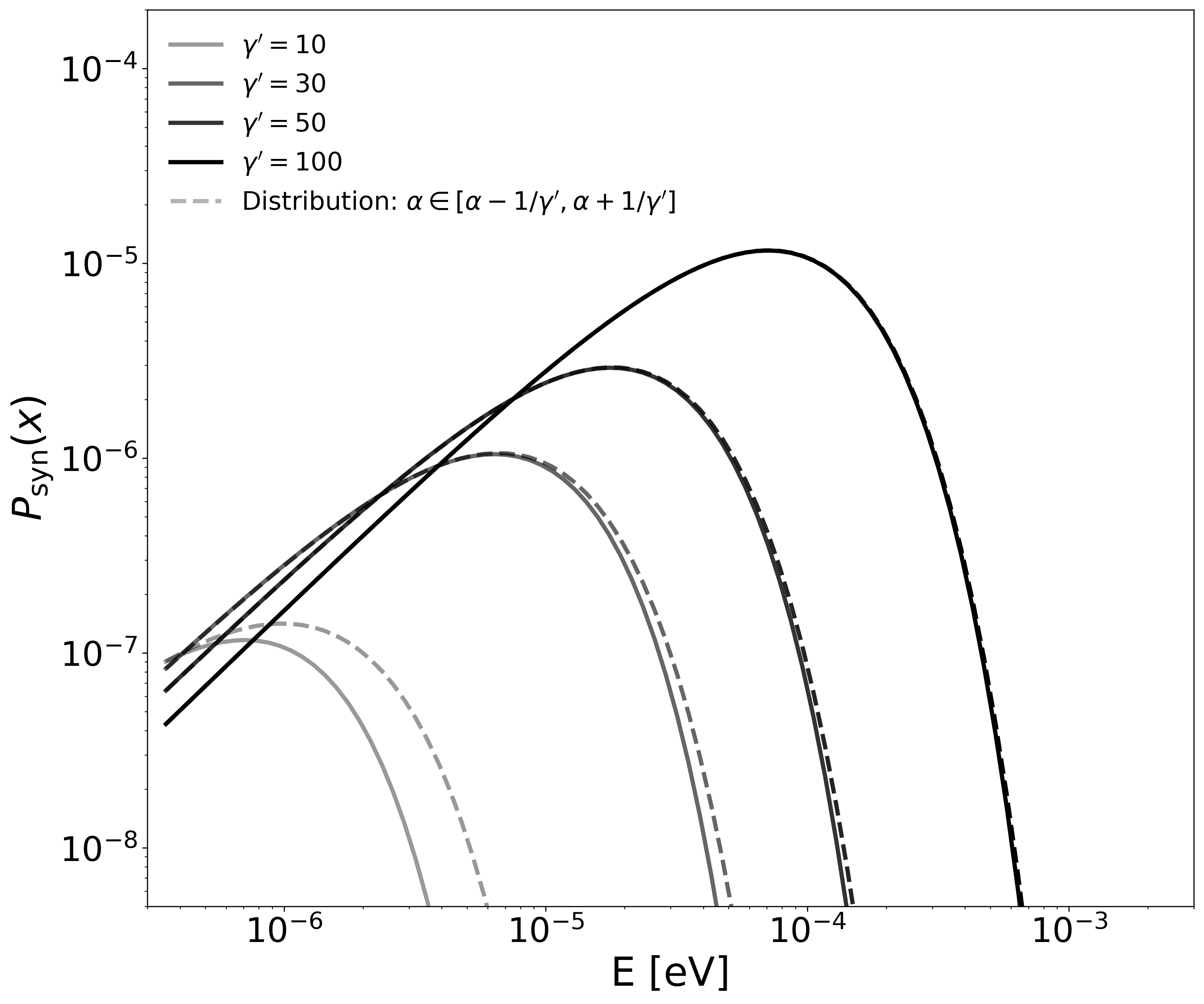}
    \includegraphics[width=1.\linewidth]{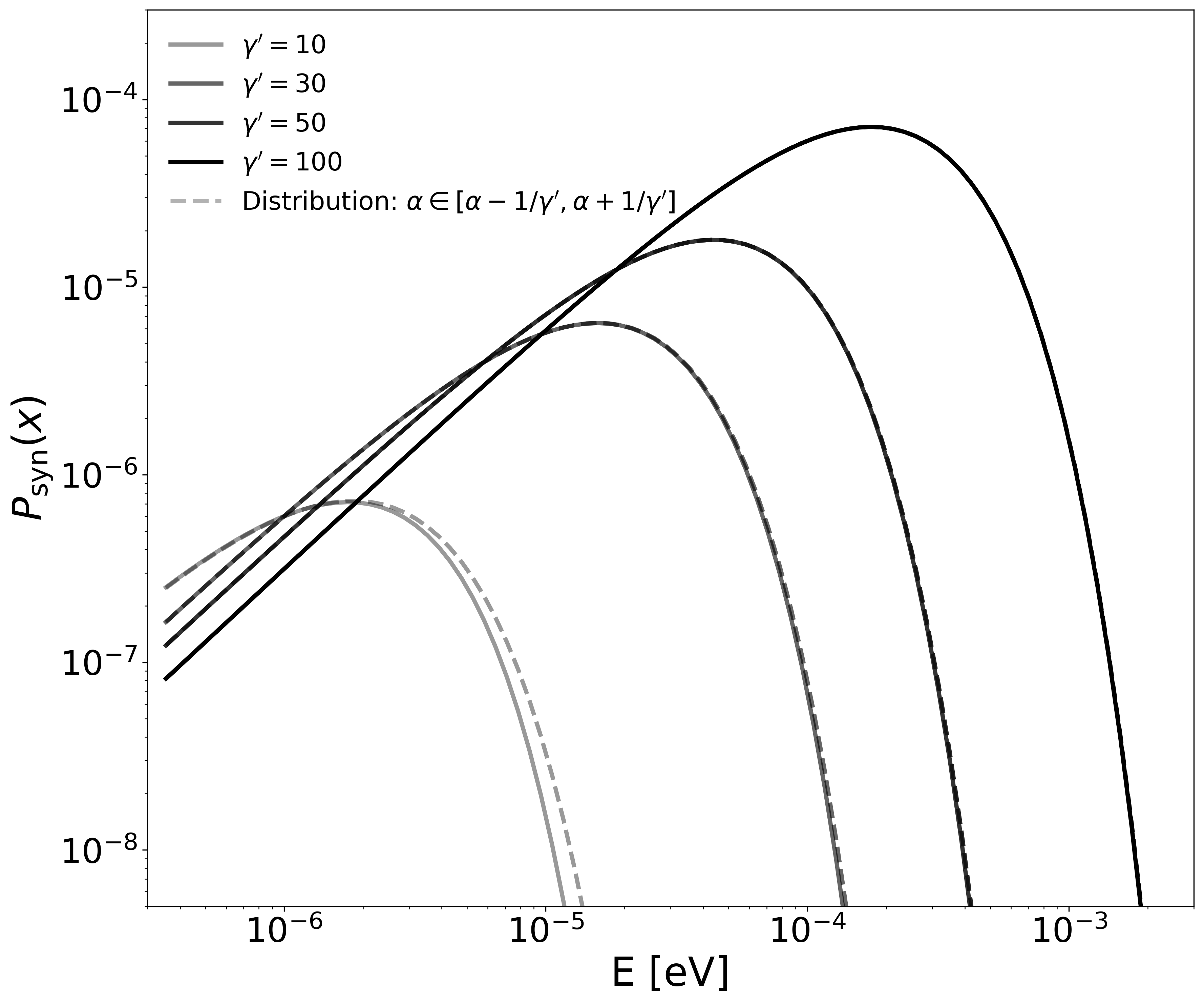}
    \caption{The synchrotron emission for fixed electron energies are presented. The dashed curves show the result if we consider electrons with different pitch angles that contribute to the emission along our LOS due to the emission cone. These are compared to the corresponding solid lines that show the synchrotron emission for one fixed pitch angle $\alpha_0$. The upper plot shows the case for $\alpha_0$ = 6$^\circ$ while the lower plot shows the results for $\alpha_0$ = 15$^\circ$.}
    \label{fig:distribution_angles_check}
\end{figure}

In the upper plot of Fig.~\ref{fig:distribution_angles_check}, the averaged synchrotron power for the distribution of pitch angles within the cone around $\alpha_0$ is compared with the synchrotron power for one fixed pitch angle $\alpha_0$ = 6$^\circ$. The dashed lines represent the synchrotron power $P'^{\mathrm{avg}}_{\mathrm{syn}}(x)$ integrated over the distribution of pitch angles within the cone. The solid lines are the synchrotron curves resulting from the fixed pitch angle. The difference between the dashed and solid curves is largest for $\gamma'_{\mathrm{e}}$ = 10,  corresponding to an electron energy of $E'_{\mathrm{e}} = 5 \cdot 10^{6}$ eV. For  $\gamma'_{\mathrm{e}}$ = 50, the difference can already be neglected.

In addition, this represents an extreme case, since it assumes that the emitted radiation is distributed uniformly across the entire beaming cone. In reality, the synchrotron emission is strongly peaked toward the centre of the cone, implying that the contribution from larger angular offsets is considerably smaller. Taking this angular dependence into account would further reduce the difference between the fixed pitch-angle approximation and the pitch-angle averaged emission.

Furthermore, blazar emission is typically modeled using a power-law electron distribution. Consequently, this effect is only relevant for the lowest-energy electrons, where the beaming cone is widest, and therefore only affects the low-energy end of the synchrotron spectrum. Even in this extreme case, a noticeable difference is found only for very small pitch angles ($\alpha \lesssim 10$$^\circ$). As demonstrated in the lower plot of Fig.~\ref{fig:distribution_angles_check}, the deviation rapidly becomes negligible with increasing pitch angle of $\alpha = 15$$^\circ$. In our simulations, we consider electrons with Lorentz factors in the range $\gamma'_{\mathrm{e}} \in [10^3,10^5]$, for which the beaming cone is orders of magnitude narrower. We therefore conclude that the effect of a finite pitch-angle distribution can safely be neglected and that the assumption of a single fixed pitch angle is justified.

\end{appendix}

\end{document}